\documentclass[aps,prb,preprint,superscriptaddress]{revtex4-1}
\usepackage{graphicx}
\usepackage{dcolumn}
\usepackage{bm}
\usepackage{amsmath}
\usepackage{xcolor}
\begin{document}



\title{Uniaxial strain-induced Kohn anomaly and electron-phonon coupling in acoustic phonons of graphene}

\author{M.E. Cifuentes-Quintal}
\email[]{cifuentes.quintal@gmail.com\\miguel.cifuentes@cinvestav.mx}
\affiliation{Departamento de F\'isica Aplicada, Centro de Investigaci\'on y de Estudios Avanzados del IPN,\\ Apartado Postal 73, Cordemex, 
97310 M\'erida, Yucat\'an, Mexico}

\author{O. de la Pe\~na-Seaman}
\affiliation{Instituto de F\'isica, Benem\'erita Universidad Aut\'onoma de Puebla,\\ Apartado postal J-48, 72570, Puebla, Puebla, Mexico}

\author{R. Heid}
\affiliation{Institut f\"ur Festk\"orperphysik, Karlsruher Institut f\"ur Technologie (KIT),\\ P.O. Box 3640, D-76021 Karlsruhe, Germany}

\author{R. de Coss}
\affiliation{Departamento de F\'isica Aplicada, Centro de Investigaci\'on y de Estudios Avanzados del IPN,\\ Apartado Postal 73, Cordemex, 
97310 M\'erida, Yucat\'an, Mexico}

\author{K.-P. Bohnen}
\affiliation{Institut f\"ur Festk\"orperphysik, Karlsruher Institut f\"ur Technologie (KIT),\\ P.O. Box 3640, D-76021 Karlsruhe, Germany}


\begin{abstract}

Recent advances in strain engineering at the nanoscale have shown the feasibility to modulate the properties of graphene. 
Although the electron-phonon (e-ph) coupling and Kohn anomalies in graphene define the phonon branches contributing to 
the resonance Raman scattering, and is relevant to the electronic and thermal transport as a scattering source, the evolution 
of the e-ph coupling as a function of strain has been less studied. 
In this work,  the Kohn anomalies and the e-ph coupling in uniaxially strained graphene along armchair (AC) and zigzag 
(ZZ) directions were studied by means of density functional perturbation theory calculations. In addition to the phonon 
anomaly at the transversal optical (TO) phonon branch in the ${\bf K}$ point for pristine graphene, we found that uniaxial 
strain induces a  discontinuity in the frequency derivative of the longitudinal acoustic (LA)  phonon branch. This behavior 
corresponds to the emergence of a Kohn anomaly, as a consequence of a strain-enhanced e-ph coupling. 
Thus, the present results for uniaxially strained graphene contrast with the  commonly assumed view that the e-ph coupling 
around the ${\bf K}$ point is only present in the TO phonon branch.

\end{abstract}


\maketitle

\section{Introduction}

After the discovery of the extraordinary properties of graphene, the next challenge is to develop  mechanisms that allow 
the enhancement and modulation of such properties. Along these lines, strain engineering is currently one of the trending topics 
in graphene science because of the possibility to induce new physical phenomena by means of mechanical strain. Examples are 
modifications on the Fermi velocity,\cite{pereira2009} the modulation of Landau levels spectra,\cite{betancur2015} the 
generation of pseudo-magnetic fields,\cite{guinea2010} the modulation of the electrical,\cite{kim2009} and thermal 
conductivities,\cite{ning2009} just to name a few. Furthermore, with the recent advances in experimental techniques to apply  
strain,  there are  different reports of uniaxial,\cite{uniax-controlado,isot-comp} biaxial,\cite{isot-comp,biax} and 
shear\cite{shear} strain in graphene. Interestingly, it has been shown that uniaxial strain can be applied in a controlled, reversible 
and non destructive way,\cite{uniax-controlado} making it of particular interest.

Two of the most studied properties of uniaxially strained graphene are its electronic and vibrational structure.
Currently, it is well know that in uniaxially strained graphene the crossing point of the valence and conduction bands at the 
Fermi level, the so called Dirac point, shifts away from the corner of the Brillouin zone (the ${\bf K}$ point), with no band 
gap opening.\cite{pereira2009,commentFarjam,Mohr2009,Choi2010} Among the vibrational structure of graphene, the $E_{2g}$ phonon
mode at the center of the Brillouin zone (the $\Gamma$ point), is particularly interesting because it is responsible 
for the G-band in the Raman spectroscopy.\cite{ferrari2007} Under uniaxial strain the $E_{2g}$ phonon mode is split in two modes, 
one parallel and the other perpendicular to the axis of the applied strain.\cite{Mohiuddin2009,cheng2011} That effect is useful to 
characterize the direction and strength of the uniaxial strain \cite{Mohiuddin2009,Huang2009,frank2013} via Raman spectroscopy. 
Even more, the full phonon dispersion,\cite{Mohr2009} Gr\"{u}neisen parameters,\cite{cheng2011} and the  origin of the phonon instability 
at the ideal strength\cite{Liu2007,Hwang2014} have been studied in uniaxially strained graphene using first principles calculations.

Regardless the level of understanding of the electronic and vibrational structure of uniaxially strained graphene,  
some fundamental and important microscopic properties like the electron-phonon (e-ph) coupling needs to be studied in detail.
In pristine graphene the e-ph coupling induces strong anomalies in the phonon dispersion,\cite{piscanec2004} contributes to the
intrinsic electronic resistivity,\cite{sohier2014} is responsible for most of the linewidth in the Raman G-band,\cite{ferrari2007} 
and determines the scattering rules for the double resonance Raman 2D-band\cite{ferrari2007}.
Even more interesting could be the possibility to induce electron-phonon superconductivity by means of atomic-decorating,\cite{profeta2012,Ludbrook2015}
heavy doping,\cite{margine2014} and a combination of doping and biaxial tensile strain.\cite{Chen-si2}
Therefore, in order to have a deep understanding of the effects of uniaxial strain on the vibrational, thermal and transport properties,  
a detailed study of the e-ph coupling in uniaxially strained graphene is mandatory.

A key feature of the e-ph coupling are Kohn anomalies:\cite{KA1} anomalous behavior in the phonon dispersion due to an electronic 
screening of the ionic vibration, which are fully determined by the geometry of the Fermi surface. In graphene the Fermi surface is the 
Dirac point, thus the Kohn anomalies take place only at the ${\Gamma}$ and ${\bf K}$ points, which are shown as a discontinuity in 
the frequency derivative of the phonon dispersion of the highest optical (HO) branches.\cite{piscanec2004} Therefore, the e-ph coupling 
is  localized on the transversal (TO) and longitudinal (LO) optical branches at ${\Gamma}$ in the $E_{2g}$ phonon mode. Meanwhile, 
at ${\bf K}$ the e-ph coupling is  almost entirely localized on the TO $A'_1$ phonon mode, with a very small contribution from 
the double degenerated $E'$ phonon mode on the LO and the longitudinal acoustic (LA) phonon branches,\cite{piscanec2004} which is usually 
neglected in the study of properties related to the e-ph coupling. In uniaxially strained graphene, the loss of hexagonal symmetry and 
the  shift of the Dirac point is expected to have an effect on the e-ph coupling but, to the best of the authors knowledge, 
this is not yet reported. 

In this work we have employed first principles density functional theory (DFT) calculations to systematically study  the effects of uniaxial 
strain along the armchair (AC) and zigzag (ZZ) directions on the Kohn anomalies and e-ph coupling in graphene. In particular, we determine 
the displacement of the Kohn anomaly from ${\bf K}$, its frequency softening, vibrational modes and e-ph coupling.
\footnote{We use the definition of e-ph coupling in graphene following the work of Piscanec et al.\cite{piscanec2004} 
The details are explained in Appendix \ref{epc-definition}.}
We show that uniaxial strain induce  a substantial enhancement of the e-ph coupling in the LA branch around ${\bf K}$ with respect to pristine
graphene, generating a non-negligible Kohn anomaly.

This paper is organized as follow: In Sec. \ref{methodology} we describe the computational details of our first principles calculations.
In Sec. \ref{res-struct} we present the changes in the bond length and average force constants that will be useful for forthcoming discussions. 
The Kohn anomalies in the phonon 
dispersion, its vibrational phonon modes and frequency shifts are shown in Sec. \ref{results-Kohn anomaly}. An analysis of the uniaxial strain 
effects in the e-ph coupling is discussed in Sec. \ref{e-ph coupling-res}. In Sec. \ref{conclusions-sec} we summarize our main findings.
Finally, we include four Appendix sections with several useful relations used along this article, concerning the structural properties of
uniaxially strained graphene, the Kohn anomaly shift, the classical atomic displacement of the discussed phonon modes, and the calculated
e-ph coupling quantities.

\section{Computational details}  \label{methodology}

The present calculations were performed within DFT, in the framework of the Mixed Basis Pseudopotential approach (MBPP).\cite{MBPP} 
Core electrons were replaced by norm conserving pseudopotentials\cite{vanderbiltPP} with non-linear core-corrections included. Valence 
states were represented by a combination of $s$ and $p$ type localized functions at each atomic sites, complemented with plane waves up to 
a kinetic energy of 25 Ry. The exchange-correlation functional was treated with the PBE\cite{PBE} parameterization of the generalized gradient 
approximation. During the structural optimization, the carbons positions were relaxed until the interatomic forces were 0.0001 Ry/Bohr or less.

For phonon and e-ph coupling calculations we employed the density functional perturbation theory as implemented in the MBPP code.\cite{Pert}
Special attention was paid to the integration in the irreducible Brillouin zone with a 72$\times$72$\times$1 Monkhorst-Pack $k$-point mesh and 
a small Gaussian broadening of 0.10 eV.
This was needed in order to avoid electronic smearing effects on the Kohn anomalies and at the same time obtain converged phonon frequencies.
Dynamical matrices were calculated using 12$\times$12$\times$1 and 9$\times$9$\times$1  $q$-points grids for pristine and 
uniaxial strained graphene, respectively. Full phonon dispersion and force constants were obtained via standard Fourier interpolation.
In order to resolve the Kohn anomalies on the phonon dispersion we also computed several low-symmetry $q$-points corresponding to the full 
$q$-grid of 72$\times$72$\times$1. For the evaluation of the e-ph coupling properties we used a denser $k$-grid of 144$\times$144$\times$1,
within a Gaussian broadening varying from 0.05 to 0.30 eV which, however, does not affect our final results.
To simulate a single atomic layer, we used the supercell approach and we left at least 12 ${\rm \AA}$ of vacuum space between successive layers 
to avoid spurious supercell effects on the electronic states and phonon frequencies.

\section{Results and Discussion}

\subsection{Structural properties} \label{res-struct}

For pristine graphene  we have obtained a lattice parameter of 2.465 \AA ,  which corresponds to a bond length of 1.423 \AA. 
Taking the derivative of the acoustic phonon branches in the limit of ${\bf q}\rightarrow 0$, we estimate a Young modulus of 369 N/m  
and a Poisson's ratio of 0.182. The calculated elastic constants values are in agreement with the previously experimental and 
computational reported values. 
For instance, Politano \textit{et al.}\cite{Politano20124903} perform phonon dispersion measurements from
macroscopic graphene samples, and estimate a Young modulus of 342 N/m and a Poisson's ratio of 0.19 from the sound velocities of 
the TA and LA phonon branches. 
In the context of previous DFT-based reports, the Young Modulus value varies from 344 to 356 N/m,\cite{Cadelano2010,Liu2007,Bera2010}
and the Poisson's ratio from 0.162 to 0.186,\cite{Cadelano2010,Gui2008,Liu2007,Bera2010} 
depending on the exchange-correlation functional and other numerical approximations.

\begin{figure}
\includegraphics*[scale=1.0]{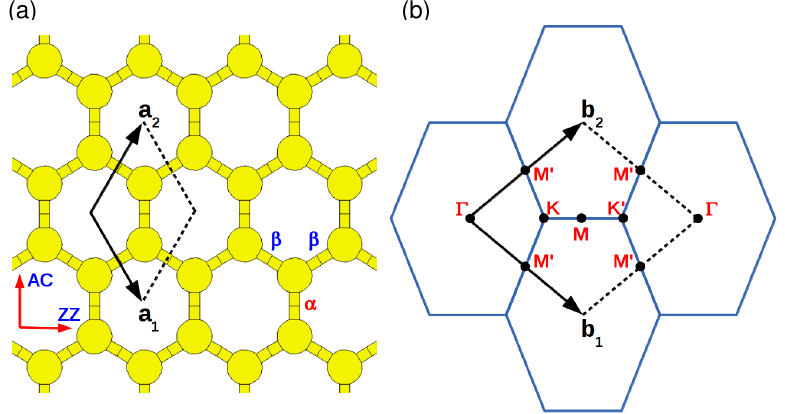}
\caption{ 
(a) Schematic representation of the lattice vectors, C-C distances ($\alpha$ and $\beta$), and strain directions (AC and ZZ) employed 
in this work. (b) First Brillouin zone in the reciprocal space with the high symmetry points for uniaxially strained graphene.}
\label{esquema}
\end{figure}

\begin{figure}
\includegraphics*[scale=1.0]{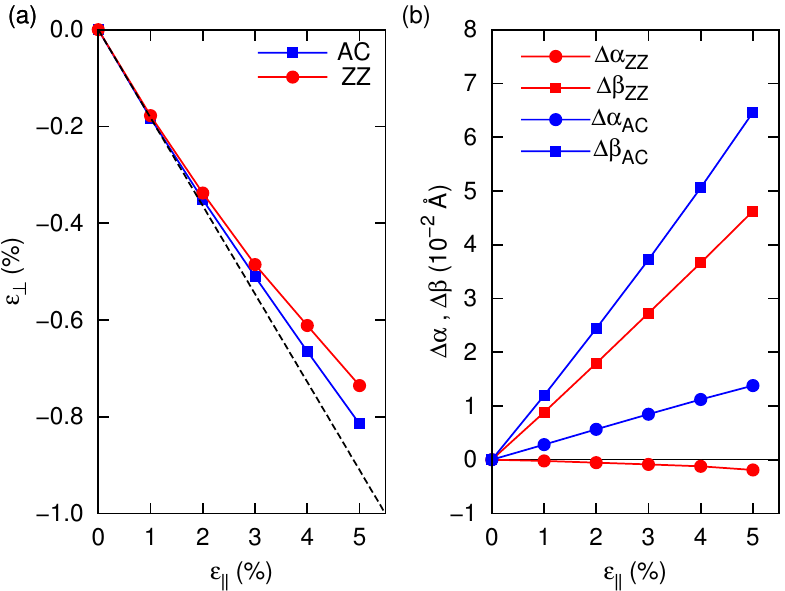}
\caption{ 
Structural properties for uniaxially strained graphene.
(a) Evolution of the perpendicular strain as a function of the parallel strain. (b) Changes of bond distances as a function of uniaxial strain.
}
\label{struct-prop}
\end{figure}

As we described in the Appendix \ref{uniax-strain}, the structural properties of graphene under ZZ and AC strain are defined by 
the relation between the parallel or applied strain $\varepsilon_{\parallel}$, the perpendicular contraction $\varepsilon_{\bot}$, 
and the C-C distances $\alpha$ and $\beta$ (see Fig. \ref{esquema} for the definition of strain directions, and the real and reciprocal lattice).
In Fig. \ref{struct-prop}(a) we present the computed values for  $\varepsilon_{\bot}$ as a function of
$\varepsilon_{\parallel}$, and for reference we have included the linear dependence for a constant Poisson's ratio. From that, it clearly shows a 
non-linear behavior for $\varepsilon_{\parallel} > 2 \%$, which indicates a non-constant Poisson's ratio, in agreement with previous works.\cite{Liu2007,wei2009,cheng2011}
Hereinafter, for simplicity $\varepsilon_{\parallel}$ will be referred only as strain.

\begin{figure}
\includegraphics*[scale=1.0]{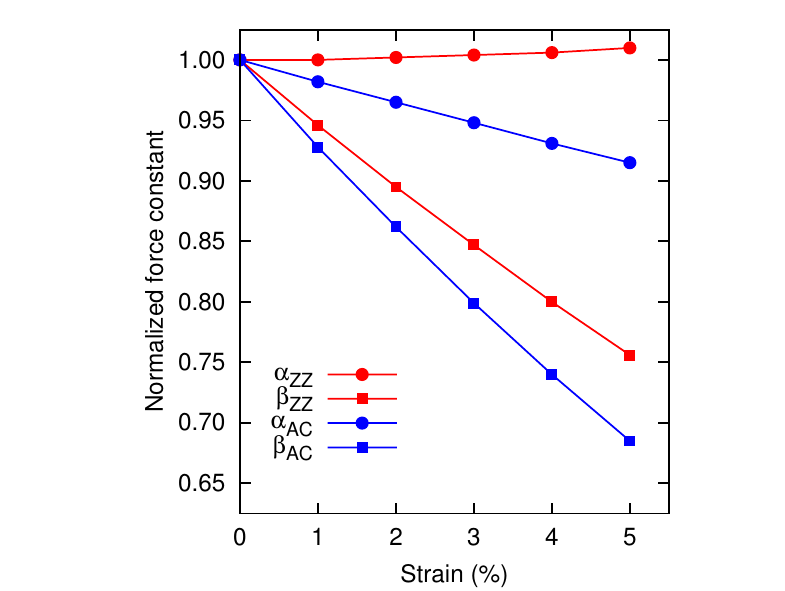}
\caption{ 
Average force constants normalized with respect to value for the bond in pristine graphene as a function of uniaxial strain.}
\label{force-ctes}
\end{figure}

The changes in the interatomic C-C distances $\alpha$ and $\beta$ are show in Fig. \ref{struct-prop}(b). Although we consider only 
tensile strain, the C-C distances do not increase in all cases. 
For ZZ strain there is a small contraction in $\alpha$, corresponding to the bond perpendicular to the direction of the applied strain.
In a classical picture, the contraction of $\alpha$  should increase the force constant related to this bond, contrary of what is 
expected when a material is under tension. To corroborate this picture, we calculate the average force constant related to atom-atom 
bonds, defined by
\begin{equation}
\label{forcecte}
 I(b) = \sqrt{\frac{1}{3}\sum_{i j}\Phi_{i j}^2 (b)},
\end{equation}
where $\Phi_{i j}(b)$ represents the force constant matrix assigned to a bond $b$. The respective $I(\alpha)$ and $I(\beta)$  are shown in 
Fig. \ref{force-ctes} for both ZZ and AC uniaxial strain.
In all cases we found that the dominant change in $I(b)$ comes from the longitudinal component of the force constant.
Just as expected from the change in the length of the C-C bonds, 
all the average force constants decrease,
except for a small hardening in $I(\alpha)$ under ZZ strain. 
Such behavior is a key feature in the forthcoming discussion of the phonon frequency shift for the Kohn anomaly.

\subsection{Kohn anomalies} \label{results-Kohn anomaly}

\begin{figure}
\includegraphics*[scale=1.0]{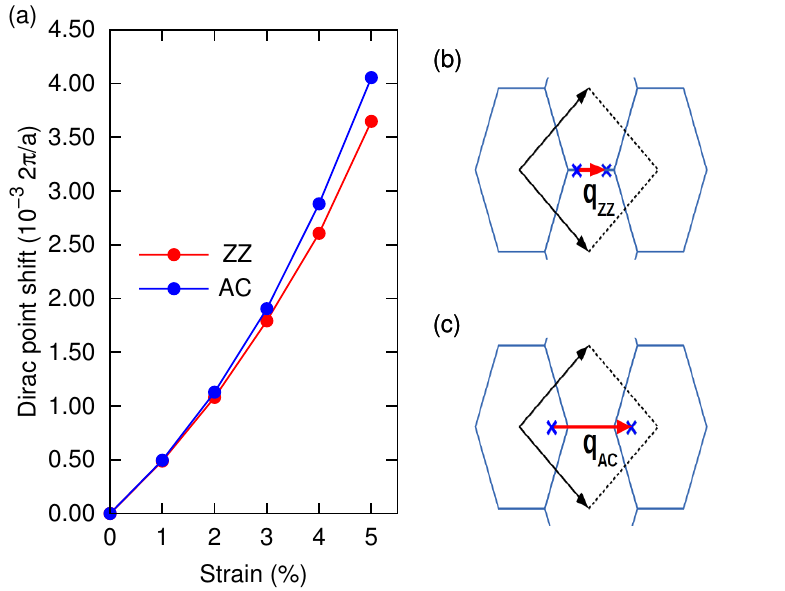}
\caption{ 
Dirac point  shift from ${\bf K}$.
(a) $\Delta$ as a function of ZZ and AC strain. Note that $a$ is the lattice parameter of  
the strained system and is given by Eq. \ref{latt-cte}. (b) Representation of ${\bf q}_{ZZ}$ and ${\bf q}_{AC}$. The cross marks represent the 
Dirac points shifted from ${\bf K}$ and ${\bf K'}$.}
\label{dp-shift}
\end{figure}

In order to determine the position of the Kohn anomalies under uniaxial strain, we need to determine the distance $\Delta$ between
the Dirac point and the ${\bf K}$ point in the electronic structure, as we described in Appendix \ref{Dirac-cone-shift}.
We estimate the evolution of $\Delta$  as a function of the applied strain by an interpolation of the electronic bands at the Fermi level 
(see Fig. \ref{dp-shift}(a)). 
We found that  $\Delta$ is bigger for strain in the AC direction than in ZZ, although for strains lower than 3 \% it is almost independent of the strain 
direction. Then, the position of the Kohn anomaly should be at the phonon nesting vector ${\bf q}_{ZZ}$ or ${\bf q}_{AC}$, 
presented for the unit cell of the reciprocal space in Figs. \ref{dp-shift}(b) and \ref{dp-shift}(c) for the ZZ and AC strain, respectively.

\begin{figure}
\includegraphics*[scale=0.6, trim=6.5cm 0cm 5cm 0cm clip]{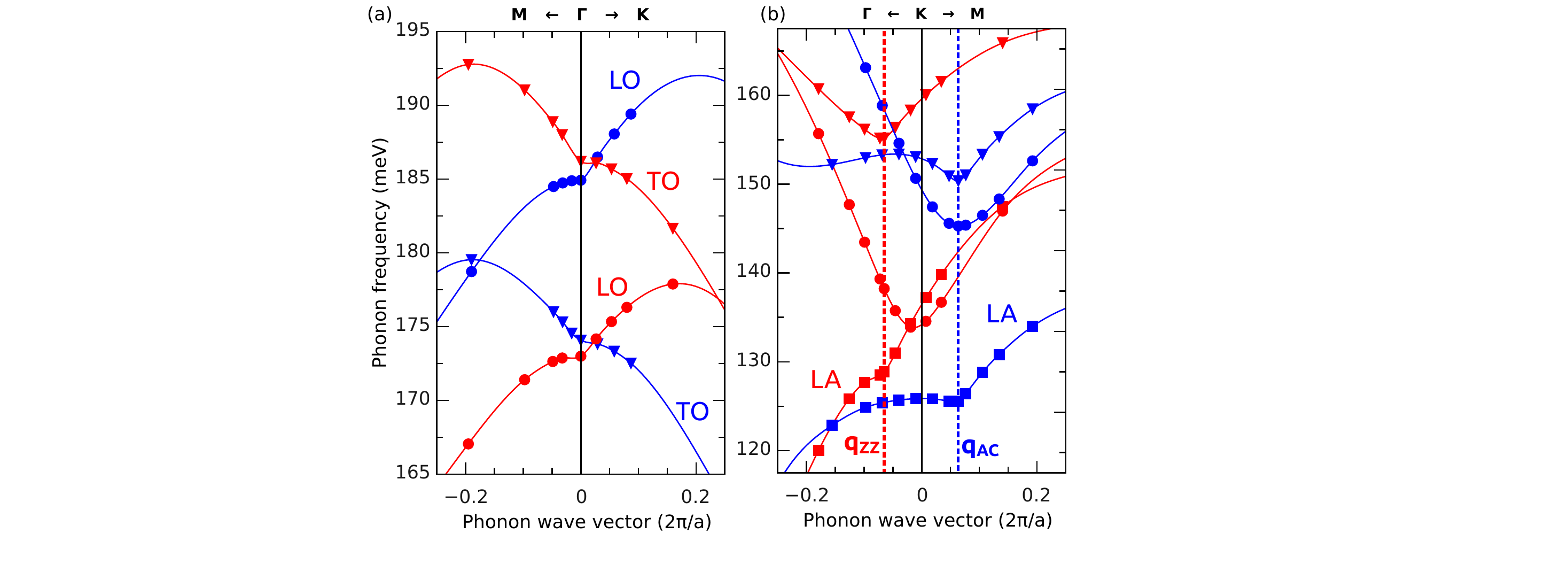}
\caption{(a) Kohn anomalies under uniaxial 5 \% of  AC (blue) and ZZ (red) strain at ${\Gamma}$, and  (b) ${\bf q}_{ZZ}$  and  ${\bf q}_{AC}$.
Dotted lines represent the position of ${\bf q}_{ZZ/AC}$. Symbols correspond to computed frequencies: circles for LO,
inverse triangles for TO, and squares for LA branches.}
\label{ph-disp}
\end{figure}

The phonon dispersion around the Kohn anomaly in uniaxially strained graphene for $\varepsilon = 5 \%$, is show in Fig. \ref{ph-disp}. 
For an easy reference and comparison, each branch and its respective phonon mode will be identified by its polarization in pristine 
graphene: LO, TO, and LA. As general trends, at ${\Gamma}$ we can observe the splitting of the $E_{2g}$ phonon mode (see Fig. 
\ref{ph-disp}(a)), and that the derivative discontinuity of the HO branches depends on the chosen direction along the Brillouin zone.
Around ${\bf K}$, the Kohn anomaly in the HO branch shows the expected shift according to our estimation for $\Delta$ (dotted line in 
Fig. \ref{ph-disp}(b)). More interesting is the new derivative discontinuity on the  LA branch at approximately 125 and 129 meV for 
the AC and ZZ strain, respectively. The fact that such discontinuities occur at the nesting vector that connect two Dirac points 
(${\bf q}_{ZZ}$ and ${\bf q}_{AC}$) is a direct indication of a Kohn anomaly in the LA branch. This is confirmed in Sec. \ref{e-ph coupling-res} 
with the analysis of e-ph coupling in the LA branch.
 
\begin{figure}
\includegraphics*[scale=1]{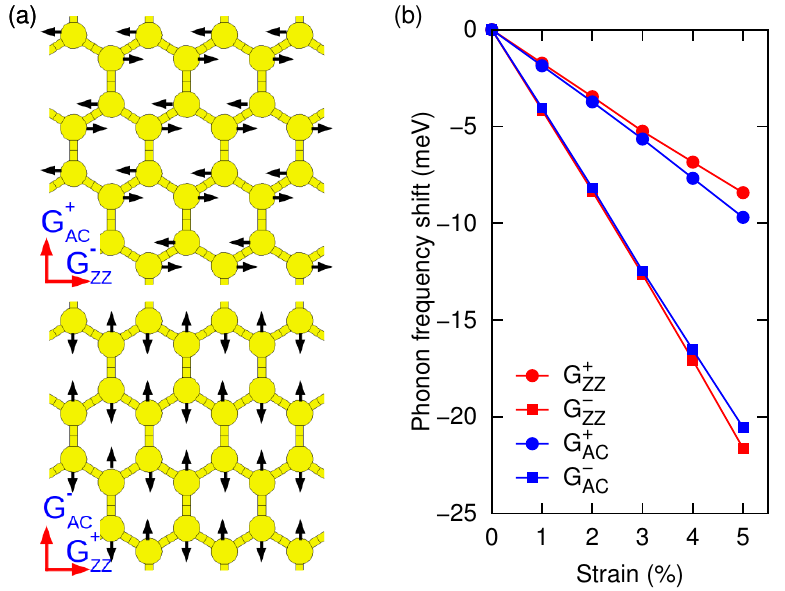}
\caption{(a) Representation of the phonon modes for ${\rm G}^+$ and ${\rm G}^-$ at ${\Gamma}$ and (b) its respective phonon frequency shift. 
The arrows indicate the instantaneous displacement of the carbon atoms at a particular time.}
\label{gamma-modes}
\end{figure}

For a further discussion of the phonon modes at the Kohn anomaly, we first focus in the ${\Gamma}$ point. 
As has been reported previously,\cite{Mohiuddin2009,cheng2011} the splitting of the $E_{2g}$ phonon mode results in two modes  with 
eigenvectors which are perpendicular (with smaller softening) and parallel to the strain direction (see Fig. \ref{gamma-modes}(a)).
This effect is measured in Raman spectroscopy via the G-band, and because its relevance in graphene characterization,
we adopt the same nomenclature that identify as ${\rm G}^+$ (${\rm G}^-$) the band with smaller (higher) softening.

The phonon frequency shift for the ${\rm G}^+$ and ${\rm G}^-$ bands are show in Fig. \ref{gamma-modes}(b).
The present results for the shift and splitting of the G-band
are in good agreement with previous theoretical\cite{Mohiuddin2009,Mohr2009,cheng2011} and experimental\cite{Mohiuddin2009} reports. 
However, it is important to mention that there are a wide range of reported values because the different set ups and conditions to 
induce strain on graphene, as well as other effects like substrate interaction, temperature, and the number of graphene layers.
Beside that, in the studied range of strain we obtain an almost linear softening in ${\rm G}^+$ and ${\rm G}^-$, which becomes 
independent of the strain direction for deformations lower than 2 \%.

\begin{figure}
\includegraphics*[scale=1]{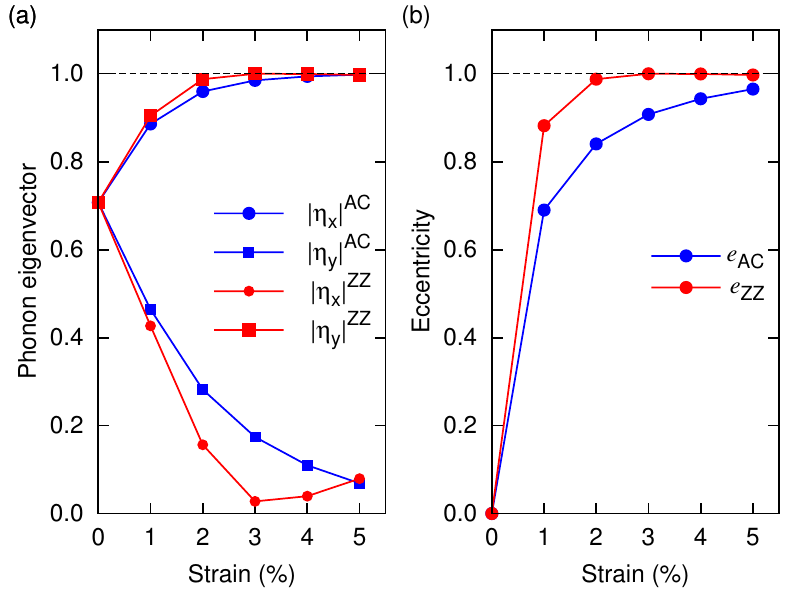}
\caption{
Polarization of the phonon eigenvectors at ${\bf q}_{ZZ}$ and ${\bf q}_{AC}$ as a function of strain. (a) Modulus of $\eta_{x,y}$ for the TO
branch, and (b) the respective eccentricity for the elliptical displacement of atoms.
}
\label{eigenvectK}
\end{figure}

For the anomalies  at  ${\bf q}_{ZZ}$ and ${\bf q}_{AC}$ in the TO and LA branches, we found a polarization of the phonon eigenvectors 
$\eta_{\kappa s}^{{\bf q}\nu}$  as a 
function of the strain, such that the atoms moves on ellipses with the mayor axis parallel (LA) and perpendicular (TO) to the strain direction, 
and whose  eccentricity approaches to one as the strain increases, until the ellipses become almost straight lines (see Appendix \ref{app-atom-disp}
for a proper description of the classical atomic displacement in graphene). 
During this evolution the phase difference $\varPhi$ between the atomic displacements along the $x$ and $y$ direction is $\pm \pi/2$,
the major and minor axes of the ellipses are defined by the magnitude of the phonon eigenvectors, and  the relations
$\left|\eta_{x}^{TO}\right| = \left|\eta_{y}^{LA}\right|$ and $\left|\eta_{y}^{TO}\right| = \left|\eta_{x}^{LA}\right|$ are always fulfilled.
This means that the Kohn anomaly shift from the high symmetry point ${\bf K}$ induces a
mixing of the phonon eigenvectors $\eta_{\kappa s}^{TO}$ and $\eta_{\kappa s}^{LA}$,
which belong to the same irreducible representation of the point group of 
${\bf q}_{ZZ}$ and ${\bf q}_{AC}$, as in pristine graphene for ${\bf q}$ points outside the high 
symmetry points $\Gamma$, ${\bf K}$, and ${\bf M}$.
Thus the classical atomic displacement on each anomaly are in mutually perpendicular ellipses,  but with the same magnitudes for the 
major and minor axes.
The magnitude of the phonon eigenvectors and the eccentricity of the resulting ellipses for the  Kohn anomaly in the TO branch are show in 
Fig. \ref{eigenvectK} with the norm $\sqrt{\left|\eta_{x}\right|^2+\left|\eta_{y}\right|^2}=1$ assumed for simplicity.
Within the ZZ (AC) strain along the $x(y)$ Cartesian axis (see Fig. \ref{esquema}), it is clear that the phonon eigenvectors 
tend to align in the strain direction, especially for ZZ strain where the eccentricity approaches one faster than for the AC strain, 
resulting in a straight line displacement.

\begin{figure}
\includegraphics*[scale=1]{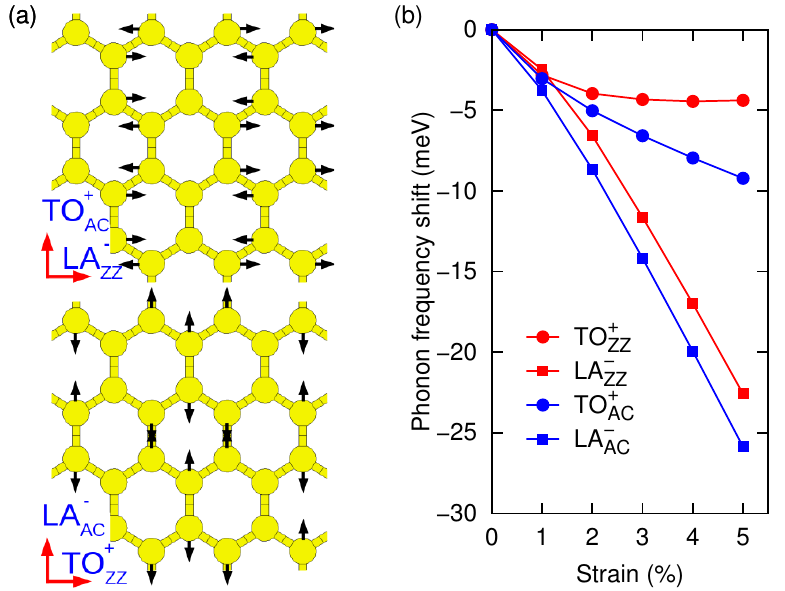}
\caption{(a) Representation of the phonon modes for the TO and LA branches at ${\bf q}_{ZZ}$ and ${\bf q}_{AC}$ and (b) the respective phonon 
frequency shift. The arrows indicate the instantaneous displacement of the carbon atoms at a particular time.}
\label{q-modes}
\end{figure}

Using the same nomenclature as that for the splitting of the G-band (${\rm G}^{-}$ and ${\rm G}^{+}$)
for the anomalies at the TO and LA branches, we will employ the $+$($-$) superindex 
to indicate that the phonon mode has eigenvectors perpendicular (parallel) to the strain direction and the smaller (higher) softening.
A schematic representation of the ${\rm TO}^{+}$ and ${\rm LA}^{-}$ modes and the behavior of the phonon frequency shift are show in 
Fig. \ref{q-modes}.
Unlike the phonon frequency shift in  ${\rm G}^{-}$, ${\rm G}^{+}$, and ${\rm LA}^{-}$, in the case of ${\rm TO}^{+}$ the phonon softening 
is non-linear and becomes nearly constant starting from  2 \% of ZZ strain. 
In  ${\rm TO}_{ZZ}^{+}$ the atoms move along the AC direction, inducing a large distortion of the $\alpha$ bond.
Thus, the constant frequency softening is a consequence of the very small increment of the force constant for the 
$\alpha$ bond, whose length remains almost constant under ZZ strain (see Fig. \ref{struct-prop} and Fig. \ref{force-ctes}).
In ${\rm TO}_{AC}^{+}$ the atoms move along the ZZ direction, the atomic distortion is not along the $\alpha$ bond,
and therefore the frequency softening is not yet constant as in  ${\rm TO}_{ZZ}^{+}$.

\subsection{Electron-phonon coupling} \label{e-ph coupling-res}

\begin{figure}
\centering
 \includegraphics*[scale=1]{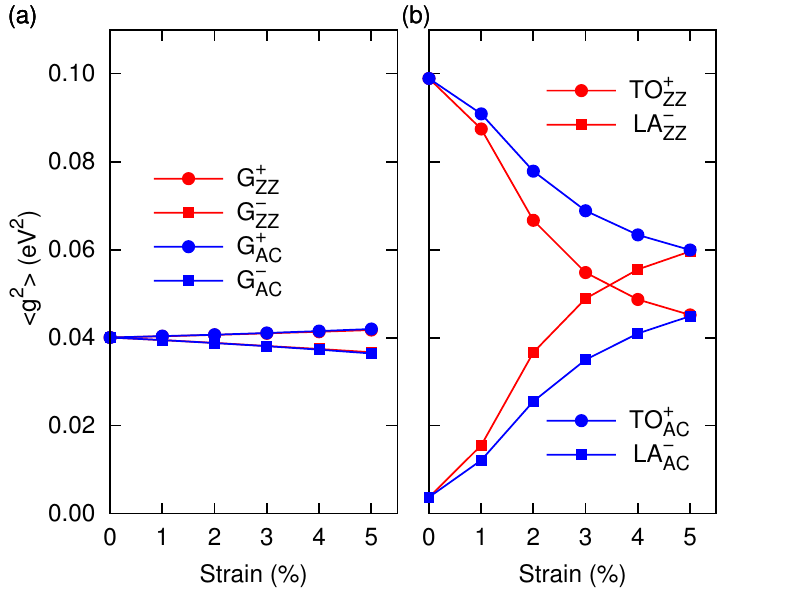}
 \caption{Electron-phonon coupling $\langle g^2\rangle $ for strained graphene  at (a) $\Gamma$ point, and (b) ${\bf q}_{ZZ}$ and ${\bf q}_{AC}$.}
\label{epc-fig}
\end{figure}

The computed values for the average e-ph coupling matrix-element square over the Fermi surface in pristine graphene are
$\langle g_{{\Gamma},G}^2\rangle = 0.0400 {\rm \ eV}^2$ and 
$\langle g_{{\bf K},TO}^2\rangle = 0.0989 {\rm \ eV}^2$, which are in excellent agreement with previously reported values.\cite{piscanec2004,Lazzeri2006, Yan2009}
We also obtain a  value of $0.0037 {\rm \ eV}^2$ for the double degenerate LO and LA branches at ${\bf K}$, which is very small in comparison with the 
TO branch.
The effect of uniaxial strain on the e-ph coupling matrix element square over the Fermi surface is shown in Fig. \ref{epc-fig}.
We report the evolution of $\langle g^2\rangle$ for ${\rm G}^+$ and ${\rm G}^-$ at the ${\Gamma}$ point (Fig. \ref{epc-fig}a), 
meanwhile for the  ${\bf q}_{ZZ}$  and ${\bf q}_{AC}$ points we analyze
the TO, LO and LA branches (Fig. \ref{epc-fig}b). In the case of the LO branch at ${\bf q}_{ZZ}$ and ${\bf q}_{AC}$ we found 
that $\langle g^2\rangle$ remains practically constant for both ZZ and AC strain, and for clarity it has not been included in Fig. \ref{epc-fig}.

At the ${\Gamma}$ point, we found that after the splitting of the $E_{2g}$ phonon mode under uniaxial strain, 
the e-ph coupling in ${\rm G}^+$ (${\rm G}^-$) slightly increases (decreases) with almost no dependence on the strain direction. 
The overall change at ${\Gamma}$ for 5\% of uniaxial strain, considering the sum of both ${\rm G}^+$ and ${\rm G}^-$, 
shows a small reduction in $\langle g_{\Gamma}^2\rangle$ of 0.0017 eV$^2$, which corresponds to -2\% with respect to the value for the pristine case.
A more complex behavior takes place in ${\bf q}_{ZZ}$ and ${\bf q}_{AC}$ as a function of uniaxial strain: the $\langle g^2\rangle$ 
in the TO branch starts to decrease with a non-linear dependence, while in the LA branch the e-ph coupling increases with almost 
the same rate as the corresponding one of the TO.
In the same way as the phonon eigenvectors polarization behave under uniaxial strain, these changes occur faster in ZZ than AC, 
with the LA branch overcoming the TO after 3\% of ZZ strain. 
However, if we take into account the sum of all the branches in ${\bf q}_{ZZ}$ and ${\bf q}_{AC}$, the total 
$\langle g_{{\bf q}}^2\rangle$ for ZZ and AC strains are very similar, and increases only by  2\% with respect to the pristine case. 
Therefore, considering both  the ${\Gamma}$ and ${\bf q}_{ZZ}$ and ${\bf q}_{AC}$ contributions to the e-ph coupling, we have that the 
total $\langle g^2\rangle$ in uniaxially strained graphene remains practically constant.

To understand the trend shown in Fig. \ref{epc-fig},
it is important to  note that $\langle g^2\rangle \sim (\delta V)^2/\omega$ (see eqs. \ref{gelements} and \ref{g2}).
Thus, considering only the contribution of $\omega$, due to the phonon softening of the Kohn anomalies, one would expected 
an increment in $\langle g^2\rangle$.
However, from Fig. \ref{epc-fig} we can see that $\langle g^2\rangle$ decreases for some phonon modes as a function of strain.
On the other hand, it should be noted that for ZZ and AC strain the pattern of the atomic vibrations in the 
$\Gamma$ point remains the same as in pristine graphene, even though the atomic distances $\alpha$ and $\beta$ are not equal. 
Meanwhile for ${\bf q}_{\rm ZZ}$ and ${\bf q}_{\rm AC}$, as a result of the Kohn anomaly shift from the high symmetry point ${\bf K}$,
the mixing of the phonon eigenvectors for the TO and LA branches induces an important 
modification of the atomic vibration with respect to pristine graphene.
Therefore, the behavior of $\langle g^2\rangle$ as function of uniaxial strain is mainly due to the change in the polarization of the phonon eigenvectors,
which contributes to the enhancement(reduction) of the e-ph coupling in the LA(TO) branch.

Regarding the anomaly  in the phonon dispersion for the LA branch at ${\bf q}_{ZZ}$ and ${\bf q}_{AC}$ discussed in Sec. \ref{results-Kohn anomaly}, 
based on the substantial increment of $\langle g^2\rangle$, it could be assigned to an emergent Kohn anomaly in uniaxially strained graphene.
This feature is a major difference in the e-ph coupling between pristine and uniaxially strained graphene, due to the presence of a new  
intervalley phonon-scattering channel for electronic states close to the Dirac point, in addition to the TO branch.
On the other hand, it was previously reported that in comparison to many-body theories which includes electronic correlation effects, 
standard DFT underestimates the e-ph coupling of the $E_{2g}$ and $A_1'$ phonon modes of graphene.\cite{lazzeri2008} 
Therefore, the inclusion of such many-body effects could give rise to a stronger Kohn anomaly and e-ph coupling than our results.
However, the use of linear response theory to compute the phonon dispersion and e-ph coupling are, at present, not implemented in 
many-body methodologies such as GW.

Here we have shown that uniaxial strain induces a non-negligible Kohn anomaly even for small strain rates, 
which opens the possibility to be experimentally observed. It would be even more interesting to evaluate 
the contribution of this anomaly to those graphene properties which depend on the e-ph coupling.
For example, it could be important to determine if this new Kohn anomaly contributes to the splitting of
the double resonance Raman scattering 2D-band, \cite{mohr2010,Yoo2D,Venezuela2D,Marzari2D,Popov2D} 
or to the intrinsic electronic resistivity,\cite{sohier2014} where until now only  the
optical $A'_1$ intervalley phonon mode was considered.

\section{Conclusions}  \label{conclusions-sec}

We have performed a first principles study of the structural properties, Kohn anomalies, and e-ph coupling for uniaxially strained graphene 
in the ZZ and AC directions. 
For ZZ strain we found a small contraction of the bond perpendicular to the strain direction, that increases the corresponding force constant.
Evaluating the shift of the Dirac point from ${\bf K}$, the phonon nesting vectors ${\bf q}_{ZZ}$ and ${\bf q}_{AC}$  were calculated.
Analyzing the phonon dispersion we found that a Kohn anomaly in ${\bf q}_{ZZ}$ and ${\bf q}_{AC}$ emerges as a function of the uniaxial
strain, in the LA branch.
For both, the original Kohn anomaly in the TO branch and the new anomaly in the LA branch, there is a polarization of the phonon eigenvectors 
as induced by the strain, in directions parallel and perpendicular to the applied strain, in the same way as is known to occur for 
${\rm G}^+$ and ${\rm G}^-$ at ${\Gamma}$. 
The softening of frequency in the Kohn anomaly shows a linear behavior as a function of the strain, except for the TO branch 
which shows a non-linear softening, and becomes almost constant for ZZ strains higher than 3\%. 
From the analysis of the average e-ph coupling matrix element square over the Fermi surface as a function of the uniaxial strain, for the
${\Gamma}$ point we found that the strain has a small effect on the ${\rm G}^+$ and ${\rm G}^-$ phonon modes. 
For the TO branch there is a reduction of $\langle g^2\rangle$ at ${\bf q}_{ZZ}$ and ${\bf q}_{AC}$,
while for the LA branch there is a large enhancement of the e-ph coupling as a function of strain.
Such behavior is mainly a consequence of the change in the polarization of the phonon eigenvectors because of mixing of the LA and TO modes induced by
the uniaxial strain.

Finally, it is important to emphasize that uniaxial strain in graphene induces a Kohn anomaly and enhancement of  the e-ph coupling in the LA 
phonon branch, in contrast with the view commonly assumed that the e-ph coupling around the ${\bf K}$ point is present only 
in the TO phonon branch.

\begin{acknowledgments}

This research was partially supported by Consejo Nacional de Ciencia y Tecnolog\'ia (CONACYT, M\'exico) under grant No. 83604.
One of the authors (M.E.C.-Q.) gratefully acknowledges a student grant from CONACYT-M\'exico, and the hospitality of the 
Karlsruher Institut f\"ur Technologie and the Benem\'erita Universidad Aut\'onoma de Puebla.

\end{acknowledgments}

\appendix
\section{Uniaxial strain} \label{uniax-strain}

In general, when an isotropic material is subjected to uniaxial mechanical strain, there is a deformation in the perpendicular direction of the 
applied strain. In the linear elastic region, the Poisson's ratio between the transverse strain ($\varepsilon_{\bot}$) and the longitudinal stain 
($\varepsilon_{\parallel}$),  defined as $-\varepsilon_{\bot} / \varepsilon_{\parallel}$, is constant. This relation could be very useful to model 
uniaxial strain, but its range of validity strongly depends on the material. In graphene, non-linear effects and relaxation of the internal atomic 
coordinates can produce a deviation from that approximation.\cite{wei2009} Therefore, in this work for a given $\varepsilon_{\parallel}$ we minimize 
the total energy as a function of $\varepsilon_{\bot}$, allowing the relaxation of the internal atomic positions in each step, in order to get vanishing 
interatomic forces.

For the description of the atomic structure let us 
consider the diatomic unit cell of graphene under two mutually perpendicular deformations. The first one is along the  AC direction and the second 
one is in the ZZ direction, as defined in Fig. \ref{esquema}(a). In such a case, the lattice vectors are given by:
\begin{eqnarray}
 \label{latt-vect}
{\bf a}_1 &=& \frac{1}{2} a_0 \left(1+\varepsilon_{ZZ} \right) {\bf \hat{x}} - \frac{\sqrt{3}}{2} a_0 \left(1+\varepsilon_{AC}\right) {\bf {\hat y}} \nonumber \\
{\bf a}_2 &=& \frac{1}{2} a_0 \left(1+\varepsilon_{ZZ} \right) {\bf \hat{x}} + \frac{\sqrt{3}}{2} a_0 \left(1+\varepsilon_{AC}\right) {\bf {\hat y}} 
\end{eqnarray} 
where $a_0$ is the lattice constant of pristine graphene, and  $\varepsilon_{AC}$ ($\varepsilon_{ZZ}$) represents the applied strain in
the AC (ZZ) direction. Under these considerations, when $\varepsilon_{\parallel}=\varepsilon_{ZZ}$ then $\varepsilon_{\bot}=\varepsilon_{AC}$, and 
\textit{vice versa}.

For both ZZ and AC strains the internal displacement of the carbon atoms is along the AC direction, with two different interatomic
distances, $\alpha$ and ${\beta}$ (see figure 1). The atomic positions in uniaxially strained graphene could be described by the relations:
\begin{eqnarray}
 \label{atom-post2}
{\bf C}_1 &=& \frac{1}{2} a_0 \left(1+\varepsilon_{ZZ} \right) {\bf \hat{x}} +  \frac{1}{2} \left( \frac{\sqrt{3}}{3} a_0 + \Delta\alpha \right) {\bf {\hat y}} \nonumber \\
{\bf C}_2 &=& \frac{1}{2} a_0 \left(1+\varepsilon_{ZZ} \right) {\bf \hat{x}} -  \frac{1}{2} \left( \frac{\sqrt{3}}{3} a_0 + \Delta\alpha \right) {\bf {\hat y}}
\end{eqnarray}
where $a_0 \sqrt{3}/3 $ is the C-C distance in pristine graphene, and $\Delta\alpha = \alpha-a_0\sqrt{3}/3$ represent the change in the
interatomic distance due to the uniaxial strain. 

\section{Kohn anomaly shift} \label{Dirac-cone-shift}

In graphene Kohn anomalies may occur only at ${\bf q}$ nesting vectors which connect two Dirac points ${\bf k}_1$ and 
${\bf k}_2 = {\bf k}_1 + {\bf q}$. Under uniaxial strain, the hexagonal symmetry of the reciprocal space is lost and a shift of the Dirac point 
from ${\bf K}$ is induced, consequently there is also  a displacement of the Kohn anomaly away from ${\bf q} = {\bf K}$. To follow such displacement, 
we employ the following reciprocal lattice vectors: 
\begin{eqnarray}
 \label{blatt-vect}
{\bf b}_1 &=& \frac{a}{a_0} \frac{1}{\left( 1+\varepsilon_{ZZ} \right)} {\bf \hat{x}} - \frac{a}{a_0} \frac{1}{\sqrt{3}\left( 1+\varepsilon_{AC} \right)} {\bf \hat{y}} \nonumber \\
{\bf b}_2 &=& \frac{a}{a_0} \frac{1}{\left( 1+\varepsilon_{ZZ} \right)} {\bf \hat{x}} + \frac{a}{a_0} \frac{1}{\sqrt{3}\left( 1+\varepsilon_{AC} \right)} {\bf \hat{y}}
\end{eqnarray}
where 
\begin{eqnarray}
 \label{latt-cte}
a &=&  \left| {\bf a}_{1,2} \right| = \frac{1}{2} a_0 \sqrt{ \left( 1 + \varepsilon_{ZZ} \right)^2 + 3 \left( 1 + \varepsilon_{AC} \right)^2 }
\end{eqnarray}
is the lattice constant, and the reciprocal space is given in units of $2\pi/a$. 

The Brillouin zone corresponding to the uniaxially strained graphene is schematically represented in Fig. \ref{esquema}(b),
with the high-symmetry points for the uniaxially strained system given by the relations:
\begin{eqnarray}
 \label{k-points}
{\bf K } &=&  \frac{2}{3} \left( \frac{a}{a_0} \right)^3 \frac{1}{\left(1+\varepsilon_{ZZ}\right)\left(1+\varepsilon_{AC}\right)^2}{\bf \hat{x}} \nonumber \\ 
{\bf M } &=&  \left( \frac{a}{a_0} \right) \frac{1}{1+\varepsilon_{ZZ}} {\bf \hat{x}}  \\
{\bf K' } &=&  2 {\bf M} - {\bf K}. \nonumber
\end{eqnarray}
For ZZ (AC) strain in accord to the Fig. \ref{esquema}, the Dirac point shifts to the left (right) of the ${\bf K}$ point. 
Considering $\Delta$ as the  distance between the Dirac point and
the ${\bf K}$ point, the nesting vectors ${\bf q}$ which indicate the position of the Kohn anomaly (formerly at ${\bf K}$), are:
\begin{eqnarray}
 \label{eq-dp-shift}
{\bf q}_{ZZ} &=& 2\left({\bf M}-{\bf K}-\Delta{\bf {\hat x}} \right) \nonumber \\
{\bf q}_{AC} &=& 2\left({\bf M}-{\bf K}+\Delta{\bf {\hat x}} \right),
\end{eqnarray}
which means that the shift of the Kohn anomaly should be along the ${\bf K}-{\bf M}$ line.

\section{Classical atomic displacement} \label{app-atom-disp}
For a given phonon mode ${\bf q}\nu$ with frequency $\omega_{{\bf q}\nu}$, the classical atomic displacement ${\bf u}_{\kappa m}^{{\bf q}\nu} $ 
as a function of the time $t$ for the $\kappa$-th atom in the $m$-th unit cell is:
\begin{equation}
 \label{u-gral}
{\bf u}_{\kappa m}^{{\bf q}\nu} = \sum_{ s} \left| \eta_{\kappa s}^{{\bf q}\nu}  \right| \cos \left( {\bf q \cdot R }_m+ \varphi_{\kappa s}^{{\bf q}\nu} -\omega_{{\bf q}\nu} t\right){\bf \hat{s}},
\end{equation}
where $\eta_{\kappa s}^{{\bf q}\nu} $ is the complex eigenvector of the phonon mode ${\bf q}\nu$ with phase $\varphi_{\kappa s}^{{\bf q}\nu}$ 
along the Cartesian  direction $s$, while ${\bf R}_m$ is the position vector of the unit cell.

In a particular unit cell of graphene,  the atomic displacement for in-plane phonon modes is reduced to
\begin{equation}
\label{u-inplane}
{\bf u}_{\kappa }^{{\bf q}\nu} = \left| \eta_{\kappa x}^{{\bf q}\nu}  \right| \cos \left( \varPhi_{\kappa}^{{\bf q}\nu} -\omega_{{\bf q}\nu} t\right){\bf \hat{x}} + 
 \left| \eta_{\kappa y}^{{\bf q}\nu}  \right| \cos \left( \omega_{{\bf q}\nu} t\right){\bf\hat{y}}, 
 \end{equation}
where $\varPhi_{\kappa}^{{\bf q}\nu}$ is the phase difference between the $x$ and the $y$ direction.

From Eq. \ref{u-inplane} we can see that each carbon atom oscillates in elliptical orbits around its equilibrium position given by 
Eq. \ref{atom-post2}. For $ \varPhi_{\kappa}^{{\bf q}\nu} = n \pi $ with $n$ an integer number, the atoms moves in straight lines with 
a slope of $ \left| \eta_{\kappa y}^{{\bf q}\nu}  \right| / \left| \eta_{\kappa x}^{{\bf q}\nu}  \right|$. In particular, when 
$\left| \eta_{\kappa y}^{{\bf q}\nu}  \right|=0$ or $\left| \eta_{\kappa x}^{{\bf q}\nu}  \right|=0$, the atoms moves respectively along the $x$ or $y$ 
axis, regardless of $ \varPhi_{\kappa}^{{\bf q}\nu}$.
If $\left| \eta_{\kappa x}^{{\bf q}\nu}  \right| = \left| \eta_{\kappa y}^{{\bf q}\nu}  \right|$ and $ \varPhi_{\kappa}^{{\bf q}\nu} = n \pi/2$ 
with $n$ an integer number different from zero, the atoms moves in circular orbits in counterclockwise for $n>0$, and clockwise for $n<0$. 

In pristine graphene, for the Kohn anomaly at ${\Gamma}$, each one of the degenerate $E_{2g}$ modes correspond to 
$\left| \eta_{\kappa x}^{{\bf q}\nu}  \right|=0$ or $\left| \eta_{\kappa y}^{{\bf q}\nu}  \right|=0$, 
meanwhile for the second anomaly at ${\bf K}$ in the $A'_1$ mode shows the conditions for circular orbits. 
Consequently, between these two anomalies along the TO branch, the atomic vibrations correspond 
to elliptical orbits whose eccentricity varies from 1 in the $E_{2g}$ mode, to 0 in the $A'_1$ mode.
This behavior of the atomic vibrations is due to a mixing of the phonon eigenvectors of the TO and LA branches, which 
belong to the same irreducible representation of the point group of ${\bf q}$ outside the high symmetry points 
${\Gamma}$, ${\bf K}$, and ${\bf M}$.

The atomic vibrations of the Kohn anomalies, induce large bond distortions that couples 
to electronic states close to the Dirac points through intravalley (${\bf q} \approx 0 $) or intervalley  (${\bf q} \approx {\bf K} $) phonon 
scattering, resulting in strong e-ph coupling.\cite{Yan2008} Therefore, modifications on the vibrational phonon mode of the Kohn anomaly 
should induce changes in the e-ph coupling.

\section{Electron-phonon coupling}\label{epc-definition}

In a metal, the strength of the e-ph coupling for a given phonon mode ${\bf q}\nu$ is characterized
by the dimensionless constant $\lambda_{{\bf q}\nu}$:
\begin{eqnarray}
\label{lambda}
\lambda_{{\bf q}\nu} &=& \frac{2}{\hbar \omega_{{\bf q}\nu} N(E_F)} \sum_{{\bf k}i j}|g_{({\bf k +q})j,{\bf k}i}^{{\bf q}\nu}|^2 \times \nonumber \\ 
& & \times \delta(\epsilon_{{\bf k}i}-E_F) \delta(\epsilon_{({\bf k+q})j}-E_F),
\end{eqnarray}
with $N(E_F)$  as the electronic density of states per atom and spin at the Fermi level $E_F$.
The e-ph coupling matrix element $g$ represents the probability of scattering from an electronic state  $\epsilon_{{\bf k}i}$ with momentum 
${\bf k}$  and band index $i$, to another state  $\epsilon_{({\bf k+q})j}$ via the absorption or emission of a phonon ${\bf q}\nu$ with 
frequency $\omega_{{\bf q}\nu}$, and is defined by
\begin{equation}
\label{gelements}
g_{({\bf k+q})j,{\bf k}i}^{{\bf q}\nu} = \sqrt{\frac{\hbar}{2\omega_{{\bf q}\nu}}}\sum_{\kappa s}\frac{1}{\sqrt{M_\kappa}}\eta_{\kappa s}^{{\bf q}\nu} \\
\langle {\bf k+q},j |\delta_{\kappa s}^{\bf q}V |{\bf k},i \rangle,
\end{equation}
where $M_\kappa$ is the mass of the $\kappa$-th atom in the unit cell, and $\delta_{\kappa s}^{\bf q}V$ denotes the first-order change in the 
total crystal potential with respect to the displacement of the atom $\kappa$ in the $s$ direction.

In graphene $N(E_F)=0$, and therefore $\lambda_{{\bf q}\nu}$ is not well defined. 
Following the work of  Piscanec et al.,\cite{piscanec2004} for graphene  we characterize the strength of the e-ph coupling in the Kohn anomalies
by means of the average e-ph coupling matrix-element square over the Fermi surface  $\langle g_{{\bf q}\nu}^2\rangle$, 
defined as:
\begin{equation}
\label{g2}
\langle g_{{\bf q}\nu}^2\rangle = \frac{\sum_{{\bf k}i j}|g_{({\bf k +q})j,{\bf k}i}^{{\bf q}\nu}|^2 \delta(\epsilon_{{\bf k}i}-E_F) \delta(\epsilon_{({\bf k+q})j}-E_F)}{\sum_{{\bf k} i j} \delta(\epsilon_{{\bf k}i}-E_F) \delta(\epsilon_{({\bf k+q})j}-E_F)},
\end{equation}
where $ \sum_{{\bf k} i j} \delta(\epsilon_{{\bf k}i}-E_F) \delta(\epsilon_{({\bf k+q})j}-E_F) $ defines the phase space.
In practice, the Dirac delta functions should be broadened for a numerical evaluation. However, the smearing of the double 
delta functions is canceled when dividing by the phase space.

In pristine graphene, the Dirac point is exactly localized at ${\bf K}$, which is commensurable with $k$-grids which are multiples of 3.  
Therefore, Eq. \ref{g2} simplifies to $\langle g_{{\bf K}}^2\rangle=\sum_{i,j}^{\pi}|g_{(2{\bf K})i,{\bf K}j}|^2/4 $, and $\langle g_{\Gamma}^2\rangle=\sum_{i,j}^{\pi}|g_{({\bf K})i,{\bf K}j}|^2/4 $,
where the sums are performed on the two degenerated $\pi$ bands at the Fermi level.\cite{piscanec2004}
In uniaxially strained graphene, due to  the shift of the Dirac point from ${\bf K}$, it is not possible to obtain an exactly commensurable $k$-grid. 
Thus, we had to use the general definition of Eq. \ref{g2} with a dense $k$-grid and a small but finite smearing. 
We verify that our results does not change in the range of 0.05 to 0.30 eV of Gaussian smearing.

\bibliography{cifuentes_graphene}

\begin{thebibliography}{45}%
\makeatletter
\providecommand \@ifxundefined [1]{%
 \@ifx{#1\undefined}
}%
\providecommand \@ifnum [1]{%
 \ifnum #1\expandafter \@firstoftwo
 \else \expandafter \@secondoftwo
 \fi
}%
\providecommand \@ifx [1]{%
 \ifx #1\expandafter \@firstoftwo
 \else \expandafter \@secondoftwo
 \fi
}%
\providecommand \natexlab [1]{#1}%
\providecommand \enquote  [1]{``#1''}%
\providecommand \bibnamefont  [1]{#1}%
\providecommand \bibfnamefont [1]{#1}%
\providecommand \citenamefont [1]{#1}%
\providecommand \href@noop [0]{\@secondoftwo}%
\providecommand \href [0]{\begingroup \@sanitize@url \@href}%
\providecommand \@href[1]{\@@startlink{#1}\@@href}%
\providecommand \@@href[1]{\endgroup#1\@@endlink}%
\providecommand \@sanitize@url [0]{\catcode `\\12\catcode `\$12\catcode
  `\&12\catcode `\#12\catcode `\^12\catcode `\_12\catcode `\%12\relax}%
\providecommand \@@startlink[1]{}%
\providecommand \@@endlink[0]{}%
\providecommand \url  [0]{\begingroup\@sanitize@url \@url }%
\providecommand \@url [1]{\endgroup\@href {#1}{\urlprefix }}%
\providecommand \urlprefix  [0]{URL }%
\providecommand \Eprint [0]{\href }%
\providecommand \doibase [0]{http://dx.doi.org/}%
\providecommand \selectlanguage [0]{\@gobble}%
\providecommand \bibinfo  [0]{\@secondoftwo}%
\providecommand \bibfield  [0]{\@secondoftwo}%
\providecommand \translation [1]{[#1]}%
\providecommand \BibitemOpen [0]{}%
\providecommand \bibitemStop [0]{}%
\providecommand \bibitemNoStop [0]{.\EOS\space}%
\providecommand \EOS [0]{\spacefactor3000\relax}%
\providecommand \BibitemShut  [1]{\csname bibitem#1\endcsname}%
\let\auto@bib@innerbib\@empty
\bibitem [{\citenamefont {Pereira}\ \emph {et~al.}(2009)\citenamefont
  {Pereira}, \citenamefont {Castro~Neto},\ and\ \citenamefont
  {Peres}}]{pereira2009}%
  \BibitemOpen
  \bibfield  {author} {\bibinfo {author} {\bibfnamefont {V.~M.}\ \bibnamefont
  {Pereira}}, \bibinfo {author} {\bibfnamefont {A.~H.}\ \bibnamefont
  {Castro~Neto}}, \ and\ \bibinfo {author} {\bibfnamefont {N.~M.~R.}\
  \bibnamefont {Peres}},\ }\href {\doibase 10.1103/PhysRevB.80.045401}
  {\bibfield  {journal} {\bibinfo  {journal} {Phys. Rev. B}\ }\textbf {\bibinfo
  {volume} {80}},\ \bibinfo {pages} {045401} (\bibinfo {year}
  {2009})}\BibitemShut {NoStop}%
\bibitem [{\citenamefont {Betancur-Ocampo}\ \emph {et~al.}(2015)\citenamefont
  {Betancur-Ocampo}, \citenamefont {Cifuentes-Quintal}, \citenamefont
  {Cordourier-Maruri},\ and\ \citenamefont {de~Coss}}]{betancur2015}%
  \BibitemOpen
  \bibfield  {author} {\bibinfo {author} {\bibfnamefont {Y.}~\bibnamefont
  {Betancur-Ocampo}}, \bibinfo {author} {\bibfnamefont {M.~E.}\ \bibnamefont
  {Cifuentes-Quintal}}, \bibinfo {author} {\bibfnamefont {G.}~\bibnamefont
  {Cordourier-Maruri}}, \ and\ \bibinfo {author} {\bibfnamefont
  {R.}~\bibnamefont {de~Coss}},\ }\href {\doibase
  http://dx.doi.org/10.1016/j.aop.2015.04.026} {\bibfield  {journal} {\bibinfo
  {journal} {Ann. Phys. (NY)}\ }\textbf {\bibinfo {volume} {359}},\ \bibinfo
  {pages} {243 } (\bibinfo {year} {2015})}\BibitemShut {NoStop}%
\bibitem [{\citenamefont {Guinea}\ \emph {et~al.}(2010)\citenamefont {Guinea},
  \citenamefont {Katsnelson},\ and\ \citenamefont {Geim}}]{guinea2010}%
  \BibitemOpen
  \bibfield  {author} {\bibinfo {author} {\bibfnamefont {F.}~\bibnamefont
  {Guinea}}, \bibinfo {author} {\bibfnamefont {M.~I.}\ \bibnamefont
  {Katsnelson}}, \ and\ \bibinfo {author} {\bibfnamefont {A.~K.}\ \bibnamefont
  {Geim}},\ }\href {http://dx.doi.org/10.1038/nphys1420} {\bibfield  {journal}
  {\bibinfo  {journal} {Nat. Phys.}\ }\textbf {\bibinfo {volume} {6}},\
  \bibinfo {pages} {30} (\bibinfo {year} {2010})}\BibitemShut {NoStop}%
\bibitem [{\citenamefont {Kim}\ \emph {et~al.}(2009)\citenamefont {Kim},
  \citenamefont {Zhao}, \citenamefont {Jang}, \citenamefont {Lee},
  \citenamefont {Kim}, \citenamefont {Kim}, \citenamefont {Ahn}, \citenamefont
  {Kim}, \citenamefont {Choi},\ and\ \citenamefont {Hong}}]{kim2009}%
  \BibitemOpen
  \bibfield  {author} {\bibinfo {author} {\bibfnamefont {K.~S.}\ \bibnamefont
  {Kim}}, \bibinfo {author} {\bibfnamefont {Y.}~\bibnamefont {Zhao}}, \bibinfo
  {author} {\bibfnamefont {H.}~\bibnamefont {Jang}}, \bibinfo {author}
  {\bibfnamefont {S.~Y.}\ \bibnamefont {Lee}}, \bibinfo {author} {\bibfnamefont
  {J.~M.}\ \bibnamefont {Kim}}, \bibinfo {author} {\bibfnamefont {K.~S.}\
  \bibnamefont {Kim}}, \bibinfo {author} {\bibfnamefont {J.}~\bibnamefont
  {Ahn}}, \bibinfo {author} {\bibfnamefont {P.}~\bibnamefont {Kim}}, \bibinfo
  {author} {\bibfnamefont {J.}~\bibnamefont {Choi}}, \ and\ \bibinfo {author}
  {\bibfnamefont {B.~H.}\ \bibnamefont {Hong}},\ }\href {\doibase
  10.1038/nature07719} {\bibfield  {journal} {\bibinfo  {journal} {Nature
  (London)}\ }\textbf {\bibinfo {volume} {457}},\ \bibinfo {pages} {706}
  (\bibinfo {year} {2009})}\BibitemShut {NoStop}%
\bibitem [{\citenamefont {Wei}\ \emph {et~al.}(2011)\citenamefont {Wei},
  \citenamefont {Lanqing}, \citenamefont {Wang},\ and\ \citenamefont
  {Zheng}}]{ning2009}%
  \BibitemOpen
  \bibfield  {author} {\bibinfo {author} {\bibfnamefont {N.}~\bibnamefont
  {Wei}}, \bibinfo {author} {\bibfnamefont {X.}~\bibnamefont {Lanqing}},
  \bibinfo {author} {\bibfnamefont {H.~Q.}\ \bibnamefont {Wang}}, \ and\
  \bibinfo {author} {\bibfnamefont {J.~C.}\ \bibnamefont {Zheng}},\ }\href
  {http://stacks.iop.org/0957-4484/22/i=10/a=105705} {\bibfield  {journal}
  {\bibinfo  {journal} {Nanotechnology}\ }\textbf {\bibinfo {volume} {22}},\
  \bibinfo {pages} {105705} (\bibinfo {year} {2011})}\BibitemShut {NoStop}%
\bibitem [{\citenamefont {P\'erez~Garza}\ \emph {et~al.}(2014)\citenamefont
  {P\'erez~Garza}, \citenamefont {Kievit}, \citenamefont {Schneider},\ and\
  \citenamefont {Staufer}}]{uniax-controlado}%
  \BibitemOpen
  \bibfield  {author} {\bibinfo {author} {\bibfnamefont {H.~H.}\ \bibnamefont
  {P\'erez~Garza}}, \bibinfo {author} {\bibfnamefont {E.~W.}\ \bibnamefont
  {Kievit}}, \bibinfo {author} {\bibfnamefont {G.~F.}\ \bibnamefont
  {Schneider}}, \ and\ \bibinfo {author} {\bibfnamefont {U.}~\bibnamefont
  {Staufer}},\ }\href {\doibase 10.1021/nl5016848} {\bibfield  {journal}
  {\bibinfo  {journal} {Nano Lett.}\ }\textbf {\bibinfo {volume} {14}},\
  \bibinfo {pages} {4107} (\bibinfo {year} {2014})}\BibitemShut {NoStop}%
\bibitem [{\citenamefont {Shioya}\ \emph {et~al.}(2014)\citenamefont {Shioya},
  \citenamefont {Craciun}, \citenamefont {Russo}, \citenamefont {Yamamoto},\
  and\ \citenamefont {Tarucha}}]{isot-comp}%
  \BibitemOpen
  \bibfield  {author} {\bibinfo {author} {\bibfnamefont {H.}~\bibnamefont
  {Shioya}}, \bibinfo {author} {\bibfnamefont {M.~F.}\ \bibnamefont {Craciun}},
  \bibinfo {author} {\bibfnamefont {S.}~\bibnamefont {Russo}}, \bibinfo
  {author} {\bibfnamefont {M.}~\bibnamefont {Yamamoto}}, \ and\ \bibinfo
  {author} {\bibfnamefont {S.}~\bibnamefont {Tarucha}},\ }\href {\doibase
  10.1021/nl403679f} {\bibfield  {journal} {\bibinfo  {journal} {Nano Lett.}\
  }\textbf {\bibinfo {volume} {14}},\ \bibinfo {pages} {1158} (\bibinfo {year}
  {2014})}\BibitemShut {NoStop}%
\bibitem [{\citenamefont {Jie}\ \emph {et~al.}(2013)\citenamefont {Jie},
  \citenamefont {Yu~Hui}, \citenamefont {Zhang}, \citenamefont {Ping~Lau},\
  and\ \citenamefont {Hao}}]{biax}%
  \BibitemOpen
  \bibfield  {author} {\bibinfo {author} {\bibfnamefont {W.}~\bibnamefont
  {Jie}}, \bibinfo {author} {\bibfnamefont {Y.}~\bibnamefont {Yu~Hui}},
  \bibinfo {author} {\bibfnamefont {Y.}~\bibnamefont {Zhang}}, \bibinfo
  {author} {\bibfnamefont {S.}~\bibnamefont {Ping~Lau}}, \ and\ \bibinfo
  {author} {\bibfnamefont {J.}~\bibnamefont {Hao}},\ }\href {\doibase
  http://dx.doi.org/10.1063/1.4809922} {\bibfield  {journal} {\bibinfo
  {journal} {Appl. Phys. Lett.}\ }\textbf {\bibinfo {volume} {102}},\ \bibinfo
  {eid} {223112} (\bibinfo {year} {2013})}\BibitemShut {NoStop}%
\bibitem [{\citenamefont {He}\ \emph {et~al.}(2014)\citenamefont {He},
  \citenamefont {Gao}, \citenamefont {Tang}, \citenamefont {Duan},
  \citenamefont {Xu}, \citenamefont {Wang}, \citenamefont {Yang}, \citenamefont
  {Ge},\ and\ \citenamefont {Shen}}]{shear}%
  \BibitemOpen
  \bibfield  {author} {\bibinfo {author} {\bibfnamefont {X.}~\bibnamefont
  {He}}, \bibinfo {author} {\bibfnamefont {L.}~\bibnamefont {Gao}}, \bibinfo
  {author} {\bibfnamefont {N.}~\bibnamefont {Tang}}, \bibinfo {author}
  {\bibfnamefont {J.}~\bibnamefont {Duan}}, \bibinfo {author} {\bibfnamefont
  {F.}~\bibnamefont {Xu}}, \bibinfo {author} {\bibfnamefont {X.}~\bibnamefont
  {Wang}}, \bibinfo {author} {\bibfnamefont {X.}~\bibnamefont {Yang}}, \bibinfo
  {author} {\bibfnamefont {W.}~\bibnamefont {Ge}}, \ and\ \bibinfo {author}
  {\bibfnamefont {B.}~\bibnamefont {Shen}},\ }\href {\doibase
  http://dx.doi.org/10.1063/1.4894082} {\bibfield  {journal} {\bibinfo
  {journal} {Appl. Phys. Lett.}\ }\textbf {\bibinfo {volume} {105}},\ \bibinfo
  {eid} {083108} (\bibinfo {year} {2014})}\BibitemShut {NoStop}%
\bibitem [{\citenamefont {Farjam}\ and\ \citenamefont
  {Rafii-Tabar}(2009)}]{commentFarjam}%
  \BibitemOpen
  \bibfield  {author} {\bibinfo {author} {\bibfnamefont {M.}~\bibnamefont
  {Farjam}}\ and\ \bibinfo {author} {\bibfnamefont {H.}~\bibnamefont
  {Rafii-Tabar}},\ }\href {\doibase 10.1103/PhysRevB.80.167401} {\bibfield
  {journal} {\bibinfo  {journal} {Phys. Rev. B}\ }\textbf {\bibinfo {volume}
  {80}},\ \bibinfo {pages} {167401} (\bibinfo {year} {2009})}\BibitemShut
  {NoStop}%
\bibitem [{\citenamefont {Mohr}\ \emph {et~al.}(2009)\citenamefont {Mohr},
  \citenamefont {Papagelis}, \citenamefont {Maultzsch},\ and\ \citenamefont
  {Thomsen}}]{Mohr2009}%
  \BibitemOpen
  \bibfield  {author} {\bibinfo {author} {\bibfnamefont {M.}~\bibnamefont
  {Mohr}}, \bibinfo {author} {\bibfnamefont {K.}~\bibnamefont {Papagelis}},
  \bibinfo {author} {\bibfnamefont {J.}~\bibnamefont {Maultzsch}}, \ and\
  \bibinfo {author} {\bibfnamefont {C.}~\bibnamefont {Thomsen}},\ }\href
  {\doibase 10.1103/PhysRevB.80.205410} {\bibfield  {journal} {\bibinfo
  {journal} {Phys. Rev. B}\ }\textbf {\bibinfo {volume} {80}},\ \bibinfo
  {pages} {205410} (\bibinfo {year} {2009})}\BibitemShut {NoStop}%
\bibitem [{\citenamefont {Choi}\ \emph {et~al.}(2010)\citenamefont {Choi},
  \citenamefont {Jhi},\ and\ \citenamefont {Son}}]{Choi2010}%
  \BibitemOpen
  \bibfield  {author} {\bibinfo {author} {\bibfnamefont {S.~M.}\ \bibnamefont
  {Choi}}, \bibinfo {author} {\bibfnamefont {S.~H.}\ \bibnamefont {Jhi}}, \
  and\ \bibinfo {author} {\bibfnamefont {Y.~W.}\ \bibnamefont {Son}},\ }\href
  {\doibase 10.1103/PhysRevB.81.081407} {\bibfield  {journal} {\bibinfo
  {journal} {Phys. Rev. B}\ }\textbf {\bibinfo {volume} {81}},\ \bibinfo
  {pages} {081407} (\bibinfo {year} {2010})}\BibitemShut {NoStop}%
\bibitem [{\citenamefont {Ferrari}(2007)}]{ferrari2007}%
  \BibitemOpen
  \bibfield  {author} {\bibinfo {author} {\bibfnamefont {A.~C.}\ \bibnamefont
  {Ferrari}},\ }\href@noop {} {\bibfield  {journal} {\bibinfo  {journal} {Solid
  State Commun.}\ }\textbf {\bibinfo {volume} {143}},\ \bibinfo {pages} {45}
  (\bibinfo {year} {2007})}\BibitemShut {NoStop}%
\bibitem [{\citenamefont {Mohiuddin}\ \emph {et~al.}(2009)\citenamefont
  {Mohiuddin}, \citenamefont {Lombardo}, \citenamefont {Nair}, \citenamefont
  {Bonetti}, \citenamefont {Savini}, \citenamefont {Jalil}, \citenamefont
  {Bonini}, \citenamefont {Basko}, \citenamefont {Galiotis}, \citenamefont
  {Marzari}, \citenamefont {Novoselov}, \citenamefont {Geim},\ and\
  \citenamefont {Ferrari}}]{Mohiuddin2009}%
  \BibitemOpen
  \bibfield  {author} {\bibinfo {author} {\bibfnamefont {T.~M.~G.}\
  \bibnamefont {Mohiuddin}}, \bibinfo {author} {\bibfnamefont {A.}~\bibnamefont
  {Lombardo}}, \bibinfo {author} {\bibfnamefont {R.~R.}\ \bibnamefont {Nair}},
  \bibinfo {author} {\bibfnamefont {A.}~\bibnamefont {Bonetti}}, \bibinfo
  {author} {\bibfnamefont {G.}~\bibnamefont {Savini}}, \bibinfo {author}
  {\bibfnamefont {R.}~\bibnamefont {Jalil}}, \bibinfo {author} {\bibfnamefont
  {N.}~\bibnamefont {Bonini}}, \bibinfo {author} {\bibfnamefont {D.~M.}\
  \bibnamefont {Basko}}, \bibinfo {author} {\bibfnamefont {C.}~\bibnamefont
  {Galiotis}}, \bibinfo {author} {\bibfnamefont {N.}~\bibnamefont {Marzari}},
  \bibinfo {author} {\bibfnamefont {K.~S.}\ \bibnamefont {Novoselov}}, \bibinfo
  {author} {\bibfnamefont {A.~K.}\ \bibnamefont {Geim}}, \ and\ \bibinfo
  {author} {\bibfnamefont {A.~C.}\ \bibnamefont {Ferrari}},\ }\href {\doibase
  10.1103/PhysRevB.79.205433} {\bibfield  {journal} {\bibinfo  {journal} {Phys.
  Rev. B}\ }\textbf {\bibinfo {volume} {79}},\ \bibinfo {pages} {205433}
  (\bibinfo {year} {2009})}\BibitemShut {NoStop}%
\bibitem [{\citenamefont {Cheng}\ \emph {et~al.}(2011)\citenamefont {Cheng},
  \citenamefont {Zhu}, \citenamefont {Huang},\ and\ \citenamefont
  {Schwingenschl{\"o}gl}}]{cheng2011}%
  \BibitemOpen
  \bibfield  {author} {\bibinfo {author} {\bibfnamefont {Y.~C.}\ \bibnamefont
  {Cheng}}, \bibinfo {author} {\bibfnamefont {Z.~Y.}\ \bibnamefont {Zhu}},
  \bibinfo {author} {\bibfnamefont {G.~S.}\ \bibnamefont {Huang}}, \ and\
  \bibinfo {author} {\bibfnamefont {U.}~\bibnamefont {Schwingenschl{\"o}gl}},\
  }\href {\doibase 10.1103/PhysRevB.83.115449} {\bibfield  {journal} {\bibinfo
  {journal} {Phys. Rev. B}\ }\textbf {\bibinfo {volume} {83}},\ \bibinfo
  {pages} {115449} (\bibinfo {year} {2011})}\BibitemShut {NoStop}%
\bibitem [{\citenamefont {Huang}\ \emph {et~al.}(2009)\citenamefont {Huang},
  \citenamefont {Yan}, \citenamefont {Chen}, \citenamefont {Song},
  \citenamefont {Heinz},\ and\ \citenamefont {Hone}}]{Huang2009}%
  \BibitemOpen
  \bibfield  {author} {\bibinfo {author} {\bibfnamefont {M.}~\bibnamefont
  {Huang}}, \bibinfo {author} {\bibfnamefont {H.}~\bibnamefont {Yan}}, \bibinfo
  {author} {\bibfnamefont {C.}~\bibnamefont {Chen}}, \bibinfo {author}
  {\bibfnamefont {D.}~\bibnamefont {Song}}, \bibinfo {author} {\bibfnamefont
  {T.~F.}\ \bibnamefont {Heinz}}, \ and\ \bibinfo {author} {\bibfnamefont
  {J.}~\bibnamefont {Hone}},\ }\href {\doibase 10.1073/pnas.0811754106}
  {\bibfield  {journal} {\bibinfo  {journal} {Proc. Natl. Acad. Sci. U. S. A.}\
  }\textbf {\bibinfo {volume} {106}},\ \bibinfo {pages} {7304} (\bibinfo {year}
  {2009})}\BibitemShut {NoStop}%
\bibitem [{\citenamefont {Frank}\ \emph {et~al.}(2011)\citenamefont {Frank},
  \citenamefont {Tsoukleri}, \citenamefont {Riaz}, \citenamefont {Papagelis},
  \citenamefont {Parthenios}, \citenamefont {Ferrari}, \citenamefont {Geim},
  \citenamefont {Novoselov},\ and\ \citenamefont {Galiotis}}]{frank2013}%
  \BibitemOpen
  \bibfield  {author} {\bibinfo {author} {\bibfnamefont {O.}~\bibnamefont
  {Frank}}, \bibinfo {author} {\bibfnamefont {G.}~\bibnamefont {Tsoukleri}},
  \bibinfo {author} {\bibfnamefont {I.}~\bibnamefont {Riaz}}, \bibinfo {author}
  {\bibfnamefont {K.}~\bibnamefont {Papagelis}}, \bibinfo {author}
  {\bibfnamefont {J.}~\bibnamefont {Parthenios}}, \bibinfo {author}
  {\bibfnamefont {A.~C.}\ \bibnamefont {Ferrari}}, \bibinfo {author}
  {\bibfnamefont {A.~K.}\ \bibnamefont {Geim}}, \bibinfo {author}
  {\bibfnamefont {K.~S.}\ \bibnamefont {Novoselov}}, \ and\ \bibinfo {author}
  {\bibfnamefont {C.}~\bibnamefont {Galiotis}},\ }\href {\doibase
  10.1038/ncomms1247} {\bibfield  {journal} {\bibinfo  {journal} {Nat.
  Commun.}\ }\textbf {\bibinfo {volume} {2}},\ \bibinfo {pages} {255} (\bibinfo
  {year} {2011})}\BibitemShut {NoStop}%
\bibitem [{\citenamefont {Liu}\ \emph {et~al.}(2007)\citenamefont {Liu},
  \citenamefont {Ming},\ and\ \citenamefont {Li}}]{Liu2007}%
  \BibitemOpen
  \bibfield  {author} {\bibinfo {author} {\bibfnamefont {F.}~\bibnamefont
  {Liu}}, \bibinfo {author} {\bibfnamefont {P.}~\bibnamefont {Ming}}, \ and\
  \bibinfo {author} {\bibfnamefont {J.}~\bibnamefont {Li}},\ }\href {\doibase
  10.1103/PhysRevB.76.064120} {\bibfield  {journal} {\bibinfo  {journal} {Phys.
  Rev. B}\ }\textbf {\bibinfo {volume} {76}},\ \bibinfo {pages} {064120}
  (\bibinfo {year} {2007})}\BibitemShut {NoStop}%
\bibitem [{\citenamefont {Hwang}\ \emph {et~al.}(2014)\citenamefont {Hwang},
  \citenamefont {Ihm}, \citenamefont {Kim},\ and\ \citenamefont
  {Cha}}]{Hwang2014}%
  \BibitemOpen
  \bibfield  {author} {\bibinfo {author} {\bibfnamefont {J.}~\bibnamefont
  {Hwang}}, \bibinfo {author} {\bibfnamefont {J.}~\bibnamefont {Ihm}}, \bibinfo
  {author} {\bibfnamefont {K.~S.}\ \bibnamefont {Kim}}, \ and\ \bibinfo
  {author} {\bibfnamefont {M.~H.}\ \bibnamefont {Cha}},\ }\href {\doibase
  http://dx.doi.org/10.1016/j.ssc.2014.09.020} {\bibfield  {journal} {\bibinfo
  {journal} {Solid State Commun.}\ }\textbf {\bibinfo {volume} {200}},\
  \bibinfo {pages} {51 } (\bibinfo {year} {2014})}\BibitemShut {NoStop}%
\bibitem [{\citenamefont {Piscanec}\ \emph {et~al.}(2004)\citenamefont
  {Piscanec}, \citenamefont {Lazzeri}, \citenamefont {Mauri}, \citenamefont
  {Ferrari},\ and\ \citenamefont {Robertson}}]{piscanec2004}%
  \BibitemOpen
  \bibfield  {author} {\bibinfo {author} {\bibfnamefont {S.}~\bibnamefont
  {Piscanec}}, \bibinfo {author} {\bibfnamefont {M.}~\bibnamefont {Lazzeri}},
  \bibinfo {author} {\bibfnamefont {F.}~\bibnamefont {Mauri}}, \bibinfo
  {author} {\bibfnamefont {A.~C.}\ \bibnamefont {Ferrari}}, \ and\ \bibinfo
  {author} {\bibfnamefont {J.}~\bibnamefont {Robertson}},\ }\href {\doibase
  10.1103/PhysRevLett.93.185503} {\bibfield  {journal} {\bibinfo  {journal}
  {Phys. Rev. Lett.}\ }\textbf {\bibinfo {volume} {93}},\ \bibinfo {pages}
  {185503} (\bibinfo {year} {2004})}\BibitemShut {NoStop}%
\bibitem [{\citenamefont {Sohier}\ \emph {et~al.}(2014)\citenamefont {Sohier},
  \citenamefont {Calandra}, \citenamefont {Park}, \citenamefont {Bonini},
  \citenamefont {Marzari},\ and\ \citenamefont {Mauri}}]{sohier2014}%
  \BibitemOpen
  \bibfield  {author} {\bibinfo {author} {\bibfnamefont {T.}~\bibnamefont
  {Sohier}}, \bibinfo {author} {\bibfnamefont {M.}~\bibnamefont {Calandra}},
  \bibinfo {author} {\bibfnamefont {C.~H.}\ \bibnamefont {Park}}, \bibinfo
  {author} {\bibfnamefont {N.}~\bibnamefont {Bonini}}, \bibinfo {author}
  {\bibfnamefont {N.}~\bibnamefont {Marzari}}, \ and\ \bibinfo {author}
  {\bibfnamefont {F.}~\bibnamefont {Mauri}},\ }\href {\doibase
  10.1103/PhysRevB.90.125414} {\bibfield  {journal} {\bibinfo  {journal} {Phys.
  Rev. B}\ }\textbf {\bibinfo {volume} {90}},\ \bibinfo {pages} {125414}
  (\bibinfo {year} {2014})}\BibitemShut {NoStop}%
\bibitem [{\citenamefont {Profeta}\ \emph {et~al.}(2012)\citenamefont
  {Profeta}, \citenamefont {Calandra},\ and\ \citenamefont
  {Mauri}}]{profeta2012}%
  \BibitemOpen
  \bibfield  {author} {\bibinfo {author} {\bibfnamefont {G.}~\bibnamefont
  {Profeta}}, \bibinfo {author} {\bibfnamefont {M.}~\bibnamefont {Calandra}}, \
  and\ \bibinfo {author} {\bibfnamefont {F.}~\bibnamefont {Mauri}},\ }\href
  {\doibase 10.1038/nphys2181} {\bibfield  {journal} {\bibinfo  {journal} {Nat.
  Phys.}\ }\textbf {\bibinfo {volume} {8}},\ \bibinfo {pages} {131} (\bibinfo
  {year} {2012})}\BibitemShut {NoStop}%
\bibitem [{\citenamefont {Ludbrook}\ \emph {et~al.}(2015)\citenamefont
  {Ludbrook}, \citenamefont {Levy}, \citenamefont {Nigge}, \citenamefont
  {Zonno}, \citenamefont {Schneider}, \citenamefont {Dvorak}, \citenamefont
  {Veenstra}, \citenamefont {Zhdanovich}, \citenamefont {Wong}, \citenamefont
  {Dosanjh}, \citenamefont {Straßer}, \citenamefont {Stöhr}, \citenamefont
  {Forti}, \citenamefont {Ast}, \citenamefont {Starke},\ and\ \citenamefont
  {Damascelli}}]{Ludbrook2015}%
  \BibitemOpen
  \bibfield  {author} {\bibinfo {author} {\bibfnamefont {B.~M.}\ \bibnamefont
  {Ludbrook}}, \bibinfo {author} {\bibfnamefont {G.}~\bibnamefont {Levy}},
  \bibinfo {author} {\bibfnamefont {P.}~\bibnamefont {Nigge}}, \bibinfo
  {author} {\bibfnamefont {M.}~\bibnamefont {Zonno}}, \bibinfo {author}
  {\bibfnamefont {M.}~\bibnamefont {Schneider}}, \bibinfo {author}
  {\bibfnamefont {D.~J.}\ \bibnamefont {Dvorak}}, \bibinfo {author}
  {\bibfnamefont {C.~N.}\ \bibnamefont {Veenstra}}, \bibinfo {author}
  {\bibfnamefont {S.}~\bibnamefont {Zhdanovich}}, \bibinfo {author}
  {\bibfnamefont {D.}~\bibnamefont {Wong}}, \bibinfo {author} {\bibfnamefont
  {P.}~\bibnamefont {Dosanjh}}, \bibinfo {author} {\bibfnamefont
  {C.}~\bibnamefont {Straßer}}, \bibinfo {author} {\bibfnamefont
  {A.}~\bibnamefont {Stöhr}}, \bibinfo {author} {\bibfnamefont
  {S.}~\bibnamefont {Forti}}, \bibinfo {author} {\bibfnamefont {C.~R.}\
  \bibnamefont {Ast}}, \bibinfo {author} {\bibfnamefont {U.}~\bibnamefont
  {Starke}}, \ and\ \bibinfo {author} {\bibfnamefont {A.}~\bibnamefont
  {Damascelli}},\ }\href {\doibase 10.1073/pnas.1510435112} {\bibfield
  {journal} {\bibinfo  {journal} {Proc. Natl. Acad. Sci. U. S. A.}\ }\textbf
  {\bibinfo {volume} {112}},\ \bibinfo {pages} {11795} (\bibinfo {year}
  {2015})}\BibitemShut {NoStop}%
\bibitem [{\citenamefont {Margine}\ and\ \citenamefont
  {Giustino}(2014)}]{margine2014}%
  \BibitemOpen
  \bibfield  {author} {\bibinfo {author} {\bibfnamefont {E.~R.}\ \bibnamefont
  {Margine}}\ and\ \bibinfo {author} {\bibfnamefont {F.}~\bibnamefont
  {Giustino}},\ }\href {\doibase 10.1103/PhysRevB.90.014518} {\bibfield
  {journal} {\bibinfo  {journal} {Phys. Rev. B}\ }\textbf {\bibinfo {volume}
  {90}},\ \bibinfo {pages} {014518} (\bibinfo {year} {2014})}\BibitemShut
  {NoStop}%
\bibitem [{\citenamefont {Si}\ \emph {et~al.}(2013)\citenamefont {Si},
  \citenamefont {Liu}, \citenamefont {Duan},\ and\ \citenamefont
  {Liu}}]{Chen-si2}%
  \BibitemOpen
  \bibfield  {author} {\bibinfo {author} {\bibfnamefont {C.}~\bibnamefont
  {Si}}, \bibinfo {author} {\bibfnamefont {Z.}~\bibnamefont {Liu}}, \bibinfo
  {author} {\bibfnamefont {W.}~\bibnamefont {Duan}}, \ and\ \bibinfo {author}
  {\bibfnamefont {F.}~\bibnamefont {Liu}},\ }\href {\doibase
  10.1103/PhysRevLett.111.196802} {\bibfield  {journal} {\bibinfo  {journal}
  {Phys. Rev. Lett.}\ }\textbf {\bibinfo {volume} {111}},\ \bibinfo {pages}
  {196802} (\bibinfo {year} {2013})}\BibitemShut {NoStop}%
\bibitem [{\citenamefont {Kohn}(1959)}]{KA1}%
  \BibitemOpen
  \bibfield  {author} {\bibinfo {author} {\bibfnamefont {W.}~\bibnamefont
  {Kohn}},\ }\href@noop {} {\bibfield  {journal} {\bibinfo  {journal} {Phys.
  Rev. Lett.}\ }\textbf {\bibinfo {volume} {2}},\ \bibinfo {pages} {393}
  (\bibinfo {year} {1959})}\BibitemShut {NoStop}%
\bibitem [{Note1()}]{Note1}%
  \BibitemOpen
  \bibinfo {note} {We use the definition of e-ph coupling in graphene following
  the work of Piscanec et al.\cite {piscanec2004} The details are explained in
  Appendix \ref {epc-definition}.}\BibitemShut {Stop}%
\bibitem [{\citenamefont {Meyer}\ \emph {et~al.}()\citenamefont {Meyer},
  \citenamefont {Els\"asser},\ and\ \citenamefont {F\"ahnle}}]{MBPP}%
  \BibitemOpen
  \bibfield  {author} {\bibinfo {author} {\bibfnamefont {B.}~\bibnamefont
  {Meyer}}, \bibinfo {author} {\bibfnamefont {C.}~\bibnamefont {Els\"asser}}, \
  and\ \bibinfo {author} {\bibfnamefont {M.}~\bibnamefont {F\"ahnle}},\
  }\href@noop {} {\bibinfo  {journal} {FORTRAN90 Program for Mixed-Basis
  Pesudopotential Calculations for Crystals, Max-Planck-Institut f\"ur
  Metallforschung, Stuttgart (unpublished)}\ }\BibitemShut {NoStop}%
\bibitem [{\citenamefont {Vanderbilt}(1985)}]{vanderbiltPP}%
  \BibitemOpen
\bibfield  {journal} {  }\bibfield  {author} {\bibinfo {author} {\bibfnamefont
  {D.}~\bibnamefont {Vanderbilt}},\ }\href {\doibase 10.1103/PhysRevB.32.8412}
  {\bibfield  {journal} {\bibinfo  {journal} {Phys. Rev. B}\ }\textbf {\bibinfo
  {volume} {32}},\ \bibinfo {pages} {8412} (\bibinfo {year}
  {1985})}\BibitemShut {NoStop}%
\bibitem [{\citenamefont {Perdew}\ \emph {et~al.}(1996)\citenamefont {Perdew},
  \citenamefont {Burke},\ and\ \citenamefont {Ernzerhof}}]{PBE}%
  \BibitemOpen
  \bibfield  {author} {\bibinfo {author} {\bibfnamefont {J.~P.}\ \bibnamefont
  {Perdew}}, \bibinfo {author} {\bibfnamefont {K.}~\bibnamefont {Burke}}, \
  and\ \bibinfo {author} {\bibfnamefont {M.}~\bibnamefont {Ernzerhof}},\ }\href
  {\doibase 10.1103/PhysRevLett.77.3865} {\bibfield  {journal} {\bibinfo
  {journal} {Phys. Rev. Lett.}\ }\textbf {\bibinfo {volume} {77}},\ \bibinfo
  {pages} {3865} (\bibinfo {year} {1996})}\BibitemShut {NoStop}%
\bibitem [{\citenamefont {Heid}\ and\ \citenamefont {Bohnen}(1999)}]{Pert}%
  \BibitemOpen
  \bibfield  {author} {\bibinfo {author} {\bibfnamefont {R.}~\bibnamefont
  {Heid}}\ and\ \bibinfo {author} {\bibfnamefont {K.~P.}\ \bibnamefont
  {Bohnen}},\ }\href {\doibase 10.1103/PhysRevB.60.R3709} {\bibfield  {journal}
  {\bibinfo  {journal} {Phys. Rev. B}\ }\textbf {\bibinfo {volume} {60}},\
  \bibinfo {pages} {R3709} (\bibinfo {year} {1999})}\BibitemShut {NoStop}%
\bibitem [{\citenamefont {Politano}\ \emph {et~al.}(2012)\citenamefont
  {Politano}, \citenamefont {Marino}, \citenamefont {Campi}, \citenamefont
  {Far{\'i}as}, \citenamefont {Miranda},\ and\ \citenamefont
  {Chiarello}}]{Politano20124903}%
  \BibitemOpen
  \bibfield  {author} {\bibinfo {author} {\bibfnamefont {A.}~\bibnamefont
  {Politano}}, \bibinfo {author} {\bibfnamefont {A.~R.}\ \bibnamefont
  {Marino}}, \bibinfo {author} {\bibfnamefont {D.}~\bibnamefont {Campi}},
  \bibinfo {author} {\bibfnamefont {D.}~\bibnamefont {Far{\'i}as}}, \bibinfo
  {author} {\bibfnamefont {R.}~\bibnamefont {Miranda}}, \ and\ \bibinfo
  {author} {\bibfnamefont {G.}~\bibnamefont {Chiarello}},\ }\href {\doibase
  10.1016/j.carbon.2012.06.019} {\bibfield  {journal} {\bibinfo  {journal}
  {Carbon}\ }\textbf {\bibinfo {volume} {50}},\ \bibinfo {pages} {4903}
  (\bibinfo {year} {2012})}\BibitemShut {NoStop}%
\bibitem [{\citenamefont {Cadelano}\ \emph {et~al.}(2010)\citenamefont
  {Cadelano}, \citenamefont {Palla}, \citenamefont {Giordano},\ and\
  \citenamefont {Colombo}}]{Cadelano2010}%
  \BibitemOpen
  \bibfield  {author} {\bibinfo {author} {\bibfnamefont {E.}~\bibnamefont
  {Cadelano}}, \bibinfo {author} {\bibfnamefont {P.~L.}\ \bibnamefont {Palla}},
  \bibinfo {author} {\bibfnamefont {S.}~\bibnamefont {Giordano}}, \ and\
  \bibinfo {author} {\bibfnamefont {L.}~\bibnamefont {Colombo}},\ }\href
  {\doibase 10.1103/PhysRevB.82.235414} {\bibfield  {journal} {\bibinfo
  {journal} {Phys. Rev. B}\ }\textbf {\bibinfo {volume} {82}},\ \bibinfo
  {pages} {235414} (\bibinfo {year} {2010})}\BibitemShut {NoStop}%
\bibitem [{\citenamefont {Bera}\ \emph {et~al.}(2010)\citenamefont {Bera},
  \citenamefont {Arnold}, \citenamefont {Evers}, \citenamefont {Narayanan},\
  and\ \citenamefont {W\"olfle}}]{Bera2010}%
  \BibitemOpen
  \bibfield  {author} {\bibinfo {author} {\bibfnamefont {S.}~\bibnamefont
  {Bera}}, \bibinfo {author} {\bibfnamefont {A.}~\bibnamefont {Arnold}},
  \bibinfo {author} {\bibfnamefont {F.}~\bibnamefont {Evers}}, \bibinfo
  {author} {\bibfnamefont {R.}~\bibnamefont {Narayanan}}, \ and\ \bibinfo
  {author} {\bibfnamefont {P.}~\bibnamefont {W\"olfle}},\ }\href {\doibase
  10.1103/PhysRevB.82.195445} {\bibfield  {journal} {\bibinfo  {journal} {Phys.
  Rev. B}\ }\textbf {\bibinfo {volume} {82}},\ \bibinfo {pages} {195445}
  (\bibinfo {year} {2010})}\BibitemShut {NoStop}%
\bibitem [{\citenamefont {Gui}\ \emph {et~al.}(2008)\citenamefont {Gui},
  \citenamefont {Li},\ and\ \citenamefont {Zhong}}]{Gui2008}%
  \BibitemOpen
  \bibfield  {author} {\bibinfo {author} {\bibfnamefont {G.}~\bibnamefont
  {Gui}}, \bibinfo {author} {\bibfnamefont {J.}~\bibnamefont {Li}}, \ and\
  \bibinfo {author} {\bibfnamefont {J.}~\bibnamefont {Zhong}},\ }\href
  {\doibase 10.1103/PhysRevB.78.075435} {\bibfield  {journal} {\bibinfo
  {journal} {Phys. Rev. B}\ }\textbf {\bibinfo {volume} {78}},\ \bibinfo
  {pages} {075435} (\bibinfo {year} {2008})}\BibitemShut {NoStop}%
\bibitem [{\citenamefont {Wei}\ \emph {et~al.}(2009)\citenamefont {Wei},
  \citenamefont {Fragneaud}, \citenamefont {Marianetti},\ and\ \citenamefont
  {Kysar}}]{wei2009}%
  \BibitemOpen
  \bibfield  {author} {\bibinfo {author} {\bibfnamefont {X.}~\bibnamefont
  {Wei}}, \bibinfo {author} {\bibfnamefont {B.}~\bibnamefont {Fragneaud}},
  \bibinfo {author} {\bibfnamefont {C.~A.}\ \bibnamefont {Marianetti}}, \ and\
  \bibinfo {author} {\bibfnamefont {J.~W.}\ \bibnamefont {Kysar}},\ }\href
  {\doibase 10.1103/PhysRevB.80.205407} {\bibfield  {journal} {\bibinfo
  {journal} {Phys. Rev. B}\ }\textbf {\bibinfo {volume} {80}},\ \bibinfo
  {pages} {205407} (\bibinfo {year} {2009})}\BibitemShut {NoStop}%
\bibitem [{\citenamefont {Lazzeri}\ \emph {et~al.}(2006)\citenamefont
  {Lazzeri}, \citenamefont {Piscanec}, \citenamefont {Mauri}, \citenamefont
  {Ferrari},\ and\ \citenamefont {Robertson}}]{Lazzeri2006}%
  \BibitemOpen
  \bibfield  {author} {\bibinfo {author} {\bibfnamefont {M.}~\bibnamefont
  {Lazzeri}}, \bibinfo {author} {\bibfnamefont {S.}~\bibnamefont {Piscanec}},
  \bibinfo {author} {\bibfnamefont {F.}~\bibnamefont {Mauri}}, \bibinfo
  {author} {\bibfnamefont {A.~C.}\ \bibnamefont {Ferrari}}, \ and\ \bibinfo
  {author} {\bibfnamefont {J.}~\bibnamefont {Robertson}},\ }\href {\doibase
  10.1103/PhysRevB.73.155426} {\bibfield  {journal} {\bibinfo  {journal} {Phys.
  Rev. B}\ }\textbf {\bibinfo {volume} {73}},\ \bibinfo {pages} {155426}
  (\bibinfo {year} {2006})}\BibitemShut {NoStop}%
\bibitem [{\citenamefont {Yan}\ \emph {et~al.}(2009)\citenamefont {Yan},
  \citenamefont {Ruan},\ and\ \citenamefont {Chou}}]{Yan2009}%
  \BibitemOpen
  \bibfield  {author} {\bibinfo {author} {\bibfnamefont {J.~A.}\ \bibnamefont
  {Yan}}, \bibinfo {author} {\bibfnamefont {W.~Y.}\ \bibnamefont {Ruan}}, \
  and\ \bibinfo {author} {\bibfnamefont {M.~Y.}\ \bibnamefont {Chou}},\ }\href
  {\doibase 10.1103/PhysRevB.79.115443} {\bibfield  {journal} {\bibinfo
  {journal} {Phys. Rev. B}\ }\textbf {\bibinfo {volume} {79}},\ \bibinfo
  {pages} {115443} (\bibinfo {year} {2009})}\BibitemShut {NoStop}%
\bibitem [{\citenamefont {Lazzeri}\ \emph {et~al.}(2008)\citenamefont
  {Lazzeri}, \citenamefont {Attaccalite}, \citenamefont {Wirtz},\ and\
  \citenamefont {Mauri}}]{lazzeri2008}%
  \BibitemOpen
  \bibfield  {author} {\bibinfo {author} {\bibfnamefont {M.}~\bibnamefont
  {Lazzeri}}, \bibinfo {author} {\bibfnamefont {C.}~\bibnamefont
  {Attaccalite}}, \bibinfo {author} {\bibfnamefont {L.}~\bibnamefont {Wirtz}},
  \ and\ \bibinfo {author} {\bibfnamefont {F.}~\bibnamefont {Mauri}},\ }\href
  {\doibase 10.1103/PhysRevB.78.081406} {\bibfield  {journal} {\bibinfo
  {journal} {Phys. Rev. B}\ }\textbf {\bibinfo {volume} {78}},\ \bibinfo
  {pages} {081406} (\bibinfo {year} {2008})}\BibitemShut {NoStop}%
\bibitem [{\citenamefont {Mohr}\ \emph {et~al.}(2010)\citenamefont {Mohr},
  \citenamefont {Maultzsch},\ and\ \citenamefont {Thomsen}}]{mohr2010}%
  \BibitemOpen
  \bibfield  {author} {\bibinfo {author} {\bibfnamefont {M.}~\bibnamefont
  {Mohr}}, \bibinfo {author} {\bibfnamefont {J.}~\bibnamefont {Maultzsch}}, \
  and\ \bibinfo {author} {\bibfnamefont {C.}~\bibnamefont {Thomsen}},\ }\href
  {\doibase 10.1103/PhysRevB.82.201409} {\bibfield  {journal} {\bibinfo
  {journal} {Phys. Rev. B}\ }\textbf {\bibinfo {volume} {82}},\ \bibinfo
  {pages} {201409} (\bibinfo {year} {2010})}\BibitemShut {NoStop}%
\bibitem [{\citenamefont {Yoon}\ \emph {et~al.}(2011)\citenamefont {Yoon},
  \citenamefont {Son},\ and\ \citenamefont {Cheong}}]{Yoo2D}%
  \BibitemOpen
  \bibfield  {author} {\bibinfo {author} {\bibfnamefont {D.}~\bibnamefont
  {Yoon}}, \bibinfo {author} {\bibfnamefont {Y.~W.}\ \bibnamefont {Son}}, \
  and\ \bibinfo {author} {\bibfnamefont {H.}~\bibnamefont {Cheong}},\ }\href
  {\doibase 10.1103/PhysRevLett.106.155502} {\bibfield  {journal} {\bibinfo
  {journal} {Phys. Rev. Lett.}\ }\textbf {\bibinfo {volume} {106}},\ \bibinfo
  {pages} {155502} (\bibinfo {year} {2011})}\BibitemShut {NoStop}%
\bibitem [{\citenamefont {Venezuela}\ \emph {et~al.}(2011)\citenamefont
  {Venezuela}, \citenamefont {Lazzeri},\ and\ \citenamefont
  {Mauri}}]{Venezuela2D}%
  \BibitemOpen
  \bibfield  {author} {\bibinfo {author} {\bibfnamefont {P.}~\bibnamefont
  {Venezuela}}, \bibinfo {author} {\bibfnamefont {M.}~\bibnamefont {Lazzeri}},
  \ and\ \bibinfo {author} {\bibfnamefont {F.}~\bibnamefont {Mauri}},\ }\href
  {\doibase 10.1103/PhysRevB.84.035433} {\bibfield  {journal} {\bibinfo
  {journal} {Phys. Rev. B}\ }\textbf {\bibinfo {volume} {84}},\ \bibinfo
  {pages} {035433} (\bibinfo {year} {2011})}\BibitemShut {NoStop}%
\bibitem [{\citenamefont {Narula}\ \emph {et~al.}(2012)\citenamefont {Narula},
  \citenamefont {Bonini}, \citenamefont {Marzari},\ and\ \citenamefont
  {Reich}}]{Marzari2D}%
  \BibitemOpen
  \bibfield  {author} {\bibinfo {author} {\bibfnamefont {R.}~\bibnamefont
  {Narula}}, \bibinfo {author} {\bibfnamefont {N.}~\bibnamefont {Bonini}},
  \bibinfo {author} {\bibfnamefont {N.}~\bibnamefont {Marzari}}, \ and\
  \bibinfo {author} {\bibfnamefont {S.}~\bibnamefont {Reich}},\ }\href
  {\doibase 10.1103/PhysRevB.85.115451} {\bibfield  {journal} {\bibinfo
  {journal} {Phys. Rev. B}\ }\textbf {\bibinfo {volume} {85}},\ \bibinfo
  {pages} {115451} (\bibinfo {year} {2012})}\BibitemShut {NoStop}%
\bibitem [{\citenamefont {Popov}\ and\ \citenamefont {Lambin}(2013)}]{Popov2D}%
  \BibitemOpen
  \bibfield  {author} {\bibinfo {author} {\bibfnamefont {V.~N.}\ \bibnamefont
  {Popov}}\ and\ \bibinfo {author} {\bibfnamefont {P.}~\bibnamefont {Lambin}},\
  }\href {\doibase 10.1103/PhysRevB.87.155425} {\bibfield  {journal} {\bibinfo
  {journal} {Phys. Rev. B}\ }\textbf {\bibinfo {volume} {87}},\ \bibinfo
  {pages} {155425} (\bibinfo {year} {2013})}\BibitemShut {NoStop}%
\bibitem [{\citenamefont {Yan}\ \emph {et~al.}(2008)\citenamefont {Yan},
  \citenamefont {Ruan},\ and\ \citenamefont {Chou}}]{Yan2008}%
  \BibitemOpen
  \bibfield  {author} {\bibinfo {author} {\bibfnamefont {J.~A.}\ \bibnamefont
  {Yan}}, \bibinfo {author} {\bibfnamefont {W.~Y.}\ \bibnamefont {Ruan}}, \
  and\ \bibinfo {author} {\bibfnamefont {M.~Y.}\ \bibnamefont {Chou}},\ }\href
  {\doibase 10.1103/PhysRevB.77.125401} {\bibfield  {journal} {\bibinfo
  {journal} {Phys. Rev. B}\ }\textbf {\bibinfo {volume} {77}},\ \bibinfo
  {pages} {125401} (\bibinfo {year} {2008})}\BibitemShut {NoStop}%
\end{thebibliography}%

\end{document}